\documentstyle[aps,prd,tighten,epsf]{revtex}
\begin{document}
\draft
\newcommand{\be}{\begin{equation}}
\newcommand{\ee}{\end{equation}}
\newcommand{\ben}{\begin{eqnarray}
\displaystyle}
\newcommand{\een}{\end{eqnarray}}
\newcommand{\la}{{\lambda}}
\newcommand{\ta}{{\tilde a}}
\newcommand{\si}{{\sigma}}
\newcommand{\th}{{\theta}}
\newcommand{\C}{{\cal C}}
\newcommand{\D}{{\cal D}}
\newcommand{\cA}{{\cal A}}
\newcommand{\cR}{{\cal R}}
\newcommand{\cO}{{\cal O}}
\newcommand{\eeo}{\cO ({1 \over E})}
\newcommand{\G}{{\cal G}}
\newcommand{\cL}{{\cal L}}
\newcommand{\T}{{\cal T}}
\newcommand{\M}{{\cal M}}
\newcommand{\p}{\partial}
\newcommand{\na}{\nabla}
\newcommand{\ssum}{\sum\limits_{i = 1}^3}
\newcommand{\dssum}{\sum\limits_{i = 1}^2}

\newcommand{\tal}{{\tilde \alpha}}
\newcommand{\tim}{{\tilde \mu}}
\newcommand{\tr}{{\tilde \rho}}
\newcommand{\tir}{{\tilde r}}
\newcommand{\rp}{r_{+}}
\newcommand{\hr}{{\hat r}}
\newcommand{\rv}{{r_{v}}}
\newcommand{\dr}{{d \over d \hr}}
\newcommand{\dR}{{d \over d R}}

\newcommand{\hhf}{{\hat \phi}}
\newcommand{\hhM}{{\hat M}}
\newcommand{\hhQ}{{\hat Q}}
\newcommand{\hht}{{\hat t}}
\newcommand{\hhr}{{\hat r}}
\newcommand{\hhS}{{\hat \Sigma}}
\newcommand{\hhD}{{\hat \Delta}}
\newcommand{\hhm}{{\hat \mu}}
\newcommand{\hro}{{\hat \rho}}
\newcommand{\hhz}{{\hat z}}
\newcommand{\tB}{{\tilde B}}
\newcommand{\hT}{\hat T}
\newcommand{\tT}{\tilde T}
\newcommand{\hC}{\hat C}
\newcommand{\ep}{\epsilon}
\newcommand{\bep}{\bar \epsilon}
\newcommand{\Ga}{\Gamma}
\newcommand{\ga}{\gamma}
\newcommand{\hth}{\hat \theta}
\title{Abelian Higgs Hair for Electrically Charged Dilaton Black Holes}
\author{ Rafa{\l} Moderski}
\address{Nicolaus Copernicus Astronomical Center \protect \\
Polish Academy of Sciences \protect \\
00-716 Warsaw, Bartycka 18, Poland \protect \\
moderski@camk.edu.pl }

\author
{Marek Rogatko}
\address{
Technical University of Lublin \protect \\
20-618 Lublin, Nadbystrzycka 40, Poland \protect \\
rogat@tytan.umcs.lublin.pl \protect \\
rogat@akropolis.pol.lublin.pl}
\date{\today}
\maketitle
\smallskip
\pacs{ 04.50.+h, 98.80.Cq.}
\bigskip
\begin{abstract}
It is argued that an electrically charged dilaton black hole can
support a long range field of a Nielsen-Olesen string.  Combining both
numerical and perturbative techniques we examine the properties of an
Abelian-Higgs vortex in the presence of the black hole under
consideration.
Allowing the black hole to approach extremality we found that all
fields of the vortex are expelled from the extreme black hole.
In the {\it thin string} limit we
obtained the metric of {\it a conical} electrically
charged dilaton black hole. The effect of the vortex can be measured
from infinity justifying its characterization as black hole hair.
\end{abstract}
\vfill
\eject
\baselineskip=18pt
\par
\section{Introduction}
{\it The no hair} conjecture of black holes was attributed to Wheeler,
who motivated by the earlier researches on the uniqueness theorems for
black holes (see for a historical account, e.g., \cite{heu}) stated
that all exteriors of black hole solutions are characterized by at
most three conserved parameters: mass, angular momentum and electric
charge. However, nowadays we are faced with the discoveries of black
hole solutions in many theories in which Einstein's equations are
coupled with self-interacting matter fields. The surprising discovery
of Bartnik and McKinnon \cite{ba} of a nontrivial particle like
structure in Einstein-Yang-Mills systems opens new realms of
nontrivial solutions to Einstein-non-Abelian-gauge systems. Black
holes can be colored \cite{ba} - \cite{CC}, support a long-range
Yang-Mills hair, but these solutions are unstable \cite{UA} - \cite{UB}.
Nevertheless they exist.
\par
Mavromatos stated \cite{ma} that, the balance between the non-Abelian
gauge field repulsion and the gravitational attraction allowed for
dressing black hole solutions by a non-trivial outside horizon
configurations. It is true not only for non-Abelian gauge fields,
obeying the Gauss law constraint, but also for scalar fields which are
not bound with the Gauss law.  In Ref.\cite{th} it was analytically
proved the existence of {\it genuine} hairy black hole solutions to
the coupled Einstein-Yang-Mills Eqns. for an $su(N)$ gauge field, for
every $N$. By a {\it genuine} $su(N)$ black hole, the authors mean a
solution which is not simply the result of embedding a smaller group in
$su(N)$.
\par
The other sort of the problems is the extension of {\it the no hair}
conjecture when the topologies of some field configurations are not
trivial. One asks if the topological defects can act as hair for black
holes. Aryal {\it et al.} \cite{vi} derived the metric which described
a Schwarzschild black hole with a cosmic string passing through
it. The extension of this problem was to show \cite{tl} the existence
of the Euclidean Einstein equations corresponding to a vortex sitting
on the horizon of a black hole. The vortex cut out a slice out of the
Schwarzschild geometry. Achucarro {\it et al.}, in Ref. \cite{gg},
presented the numerical and analytical evidence that the Abelian Higgs
vortex is not just dressing for the Schwarzschild black hole, but the
vortex is truly hair, property of which can be detected by asymptotic
observers. They argued, that the Vilenkin's solution \cite{vi} is {\it
the thin vortex} limit of the physical vortex- black hole
configuration.  Chamblin {\it et al.} \cite{cha} considered the
problem of an Abelian Higgs vortex in a Reissner-Nordstr\"om
background. They found that in the extreme limit all of the fields
connected with vortex are expelled from the black hole.  They showed
numerically the evidence that the magnetic field wrapped around the
horizon.
\par
Superstring theories are taken into account as consistent quantum
theories of gravity. Numerical studies of the solutions to the
low-energy string theory, i.e., the Einstein-dilaton black holes in
the presence of a Gauss-Bonnet type term, revealed that they were
endowed with a nontrivial dilaton hair \cite{ka}. This dilaton hair is
expressed in term of its ADM mass \cite{adm}.  The extended moduli and
dilaton hair and their associated axions for a Kerr-Newmann black hole
background were computed in Ref.\cite{ka1}. The problem of an Abelian
Higgs vortex in the background of an Euclidean electrically charged
black hole was studied in Ref.\cite{mr}. It was shown that this kind
of the Euclidean black hole can support a vortex solution on a horizon
of the black hole. The vortex effect was to cut out a slice out of the
considered black hole geometry.
\par
In this paper, we shall try to provide some continuity with our
previous work \cite{mr}. Now, we shall consider the problem of an
Abelian Higgs vortex in an electrically charged dilaton black hole
background. In our considerations we allow for the presence of
two $U(1)$ gauge fields, one is originated from the low-energy string
action, the other is the symmetry spontaneously broken in the ground
state. We shall also be interested in finding the metric for the
dilaton black hole-vortex configuration, in {\it the thin string}
limit.
\par
The outline of the paper is as follows. Section II is devoted to the
general analytic considerations of the Nielsen-Olesen vortex in the
presence of an electrically charged dilaton black hole. In Sec.III, we
presented the numerical analysis of the problem. In Sec.IV, using the
iterative method of solving equations and assuming {\it the thin
string} limit, we find the metric for an electrically charged dilaton
black hole with a cosmic string passing through it. In Sec.V we finish
with general conclusions concerning our investigations.

\section{The Nielsen-Olesen vortex in the presence of an 
electrically charged dilaton black hole} 
In this section we shall study the Nielsen-Olesen Eqns. for an
Abelian Higgs vortex in the presence of an electrically charged
dilaton black hole background.
In our considerations, we assume the black hole-string vortex
configuration which involves the separation between the degrees of
freedom of each of the considered objects. Namely, the system is
described by the action
\be
S = S_{1} + S_{2},
\ee
where $S_{1}$ is the low-energy string action, see e.g., \cite{gro}
\be
S_{1} = \int \sqrt{-g}d^4 x
\left [ {R\over 16 \pi G} - 2 ( \na \phi )^2 - e^{-2 \phi} F^2 \right
], 
\ee
where $F_{\alpha \beta} = 2\na_{[ \alpha}A_{\beta ]}$, $\phi$ is the
dilaton field and $S_{2} $ is the action for an Abelian Higgs system
minimally coupled to gravity and be subject to spontaneous symmetry
breaking. It has the form
\be
S_{2} = \int \sqrt{-g}d^4 x
\left [ - D_{\mu} \Phi^{\dagger} D^{\mu} \Phi - {1 \over 4}
\tB_{\mu \nu} \tB^{\mu \nu} - {\la \over 4}
\big ( \Phi^{\dagger} \Phi - \eta^2 \big )^{2} \right ],
\ee
where $\Phi$ is a complex scalar field, $D_{\mu} = \na_{\mu} + ie
B_{\mu}$ is the gauge covariant derivative. $\tB_{\mu \nu}$ is the
field strength associated with $B_{\mu}$.  One can define the real
fields $X, P_{\mu}, \chi$ by the relations
\cite{gg}
\ben
\Phi (x^{\alpha}) = \eta X (x^{\alpha}) e^{i \chi (x^{\alpha})},\\ 
B_{\mu}(x^{\alpha}) = {1 \over e} \left [ P_{\mu}(x^{\alpha}) -
\na_{\mu} \chi (x^{\alpha}) \right ].
\een
Thus, the equations of motion derived from the action $S_{2}$ are as
follows:
\ben
\na_{\mu} \na^{\mu} X - P_{\mu}P^{\mu} X - {\la \eta^2 \over 2}
\big ( X^2 - 1 ) X = 0, \\
\na_{\mu} \tB^{\mu \nu} - 2 e^2 \eta^2 X^2 P^{\nu} = 0. 
\een
It turns out, that the field $\chi$ is not dynamical. The vortices of
the Nielsen-Olesen type, in the flat spacetime, have the cylindrically
symmetric solutions of the forms
\ben
\Phi = X (\tr) e^{i N \phi}, \\
P_{\phi} = N P(\tr),
\een
where $\tr$ is the cylinder radial coordinate , $N$ is the winding
number. 
\par
On the other hand, static, spherically symmetric solution of the Eqns.
of motion derived from the action $S_{1}$ determines the metric of an
electrically charged dilaton black hole. It is given by
\cite{db}
\be
ds^2 = - \left ( 1 - {2 GM \over \tir} \right ) dt^2 +
{ d \tir^2 \over \left ( 1 - {2 GM \over \tir} \right ) } + \tir \left ( 
\tir - {Q^2
\over GM} \right ) (d \theta^2 + \sin^2 \theta  d \phi^2).
\ee
One can define $\tir_{+} = 2GM$ and $\tir_{-} = {Q^2 \over GM}$ which
are related to the mass $M$ and charge $Q$ by the relation $Q^2 =
{\tir_{+} \tir_{-} \over 2} e^{2 \phi_{0}}$. The charge of the dilaton
black hole $Q$, couples to the field $F_{\alpha \beta}$, is unrelated
to the Abelian gauge field $B_{\mu \nu}$ associated with the
vortex. The dilaton field is given by $e^{2\phi} = \left ( 1 -
{\tir_{-} \over r} \right ) e^{-2\phi_{0}}$, where $\phi_{0}$ is the
dilaton's value at $r \rightarrow \infty$.  The event horizon is
located at $\tir = \tir_{+}$. For $\tir = \tir_{-}$ is another
singularity, one can however ignore it because $\tir_{-} <
\tir_{+}$. The extremal black hole occurs when $\tir_{-} = \tir_{+}$,
when $Q^2 = 2M^2 e^{2\phi_{0}}$.
\par
It is convenient to work with the rescaled radial coordinates and the
black hole parameters by the Higgs wavelength, i.e., with the
non-dimensional variables
\be
(r, E, q) = \sqrt{\la} \eta (\tir, GM, Q),
\ee
where $\eta$ parameter is the energy scale of symmetry breaking, $\la$
is the Higgs coupling. These parameters can be related to the Higgs
mass $m_{Higgs} = \sqrt{\la} \eta$ and the mass of the vector field in
the broken phase $m_{vect} = \sqrt{2}e \eta$.  Returning to the
Eqs. of motion, we can consistently assume that
\ben
X = X(r, \theta),\\
P_{\phi} = N P(r, \theta).
\een
Then, Eqs. of motion yield
\ben \label{m}
{1 \over r (r - {q^2 \over E})} \p_{r} \left [
\big ( r - {q^2 \over E} \big ) (r - 2 E ) \p_{r} X \right ]
&+& {1 \over  r (r - {q^2 \over E}) \sin \theta }
\p_{\theta} [ \sin \theta \p_{\theta} X ] -
{N P^2 X \over  r (r - {q^2 \over E}) \sin^2 \theta} -
{1 \over 2} X (X^2 -1 ) = 0,\\
\p_{r} \left [ \left ( 1 - {2 E \over r} \right ) \p_{r} P \right ] &+&
{\sin \theta \over  r (r - {q^2 \over E})} \p_{\theta}
[csc \theta \p_{\theta} P ] - {X^2 P \over \beta} = 0,
\label{mm}
\een
where $\beta = {\la \over 2 e^2} = m_{Higgs}^2 / m_{vect}^2$ is the
Bogomolnyi parameter.  When $ \beta \rightarrow \infty$, the Higgs
field decouples and like in the Reissner-Nordstr\"om case \cite{cha},
one can study a free Maxwell field in the electrically charged black
hole spacetime. The other situation will arise when $P = 1$. It will
be the case of a global string in the presence of the electrically
charged dilaton black hole.
\par
Noting as in Refs. \cite{gg} and \cite{gh}, that in normal spherically
symmetric coordinates $X$ is a function of $\sqrt{g_{33}}$, we
shall try with the coordinates $R = \sqrt{r (r - {q^2 \over E})} \sin
\theta$, namely $X = X(R)$ and $P_{\phi} = P_{\phi}(R)$.  Using the
$r$-coordinates and taking into account {\it the thin string} limit,
i.e., $E \gg 1$, Eqs. (\ref{m}) and (\ref{mm}) can be rewritten
\ben
{( r + \eeo ) \sin^2 \theta \over ( r^2 + \eeo ) R} 
\dR \left [ R ( r - 2E) + \eeo \right ] \dR X + {1 \over R} \dR
\left [ r \cos^2 \theta \dR X \right ] - {P^2 X N \over R^2} -
{1 \over 2} (X^2 - 1) X = \\ 
= {X' \over R} + X'' - {P^2 X N \over R^2} - {1 \over 2} (X^2 - 1) X
- {2 E \over r} \sin^2 \theta \left [ {X' \over R} + X''
\right ] = 0, \\ \nonumber
{( r + \eeo ) R \over r ( r + \eeo)} \dR
\left [ \left ( 1 - {2E \over r} \right ) {\sin^2 \theta ( r + \eeo ) \over
R} \dR P \right ] + 
R \cos^2 \theta \dR  \left [ {P' \over R} \right ]
- {X^2 P \over \beta} = \\ \nonumber
= - {P' \over R} + P'' - {X^2 P \over \beta} + {2E \sin^2 \theta \over
r} \left [ {P' \over R} - P'' \right ] = 0,
\een
where $'$ is the derivative $\dR$. The above Eqns. for $X$ and $P$ are
of the Nielsen-Olesen type, up to the errors of the order ${2E \sin^2
\theta \over r}$ multiplied by the other terms in the adequate
Eqns. Since $R = \sqrt{r (r - {q^2 \over E})} \sin \theta \sim \cO
(1)$ in the core of the string, $\sin \theta = \cO ({1 \over r})$, the
errors are of the order of $\cO ({E \over r^3}) < \cO ({1 \over E^2})
\ll 1$.  Hence, the Nielsen-Olesen vortex can sit on the electrically
charged dilaton black hole. Of course, the above considerations have
only approximate forms and do not prove the existence of a solution to
the problem. Nevertheless, they substitute a good approximation.
\par
The Eqns. of motion are rather intractable in the exact form, so one
can use an approximate method of solving them. Now, we try to describe
an analytic solution to these Eqns. for the black hole which size is
small comparing to the vortex size. One will require the limit of
$\sqrt{N} \gg E$, using the radius of a flux tube of the order $r \sim
\sqrt{2N} \beta ^{1 \over 4}$ for $N \gg 1$. The large $N$-limit was first
considered in Ref.\cite{nn}.
\par
Inside a core of the vortex the gauge symmetry will be unbroken, then
one can expect that ${X^2 \over \beta} \approx 0$. Taking into
consideration Eqn.  (\ref{mm}), we can see that the solution is
provided by
\be
P \approx 1 - p R \sin^2 \theta,
\label{p}
\ee
where $p$ is an integration constant equal to twice the magnetic field
strength at the center of the core \cite{gg}.  Using the analysis
performed in Ref.\cite{nn}, one can show that far from the
electrically charged dilaton black hole but still inside the vortex $p
\approx {1 \over 2N \sqrt{\beta}}$. Then, large $N$ means small $p$.
Following the method used in \cite{gg} and \cite{nn}, we 
set $X = \xi^N$ and try to solve Eqns.
for the Higgs fields $X$, expanding in powers of ${1 \over N}$ and
taking the limit $\cO \left ( {1 \over N^2} \right )$. Then, one has
\be
\left ( 1 - {2E \over r} \right ) \left ( {\p_{r} \xi \over \xi} \right
)^2 +
{1 \over r (r - {q^2 \over E})} \left ( {\p_{\theta} \xi \over \xi} \right
)^2 - {P^2 \over r (r - {q^2 \over E}) \sin^2 \theta} +
\cO \left ( {1 \over N^2} \right ) = 0.
\ee
Assuming that, $ \xi = b(r) \sin \theta$, one gets the following Eqn.
for $b(r)$:
\be
{1 \over b} {db \over dr}
= {1 - pr (r - {q^2 \over E}) \over \sqrt{ (r - 2E)(r -
{q^2 \over E})}} ,
\ee
which is solved by the following expression:
\ben \label{b}
b(r) = k \left [
r - \left ( E + {q^2 \over E} \right ) + \sqrt{ (r - 2 E)(
r - {q^2 \over E})}
\right ]^{\left [ 1 -{p \over 8} \left ( 6E - {q^2 \over E} \right )
\left ( 2E + {q^2 \over E} \right ) + p q^2
\right ]} \\ \nonumber
e^{ \left [
{- p \over 2}\left ( r + {1 \over 2} \left ( 6E - {q^2 \over E} \right
) \right ) \sqrt{(r - 2E)( r - {q^2 \over E})}
\right ]},
\een
where $k$ is an integration constant. We can now establish the following: 
\be
X \approx b
^{N} (r) \sin^{N} \theta.
\ee
Eqns. (\ref{p}) and (\ref{b}) consist the approximate solutions for the
case of an electrically charged dilaton black hole that sits inside the
vortex core. The thickness of the vortex, i.e., the distance at which
$P \approx 0,$
is roughly defined as
\be
r \sin \theta = {q^2 \over 2E} + {1 \over \sqrt{p}} \left (
1 +  {pq^4 \over 8 E^2} \right ).
\ee
Comparing this result to the outcome obtained in the case of the 
Abelian Higgs vortex in the Schwarzschild background \cite{gg}, one
can see that
the vortex is thicker on the equatorial region when we consider the
electrically charged dilaton black hole. The result is similar to the 
Reissner-Nordstr\"om-vortex case, where in Ref.\cite{cha}, the very similar
conclusion was reached. Calculating the magnetic flux crossing the
horizon of an electrically charged dilaton black hole
\be
\tB_{\theta \phi} \mid_{ \rp} = - p \rp \left ( \rp - {q^2 \over E}
\right ) \sin^{2} \theta,
\ee
we see that it decreases when the charge increases. For the extreme
electrically dilaton black hole, i.e., $\rp = r_{-}$, one has the
expelling of Higgs fields from the horizon. In the extreme
Reissner-Nordstr\"om black hole case Chamblin {\it et al.}
get the similar 
result, see Ref.\cite{cha}.
Both these results are of $p \sim {1 \over N}$ order and are relatively
small due to the small value of the charge $q$ comparing to  $E$.
\section{Numerical solutions}
In this section we shall analyze Eqns. (\ref{m}) and (\ref{mm})
numerically, outside and on the horizon of the black hole.  The
Eqns. under considerations are elliptic in the electrically charged
dilaton black hole background, while on the horizon of the black hole
they are parabolic. At large radii, one wants to obtain the
Nielsen-Olesen solutions, then the asymptotic values of the functions
$X$ and $P$ are
\be (X, P) = \cases{ (1, 0), & $r \rightarrow \infty$
\cr (0, 1),& $r \ge 2E,  \theta = 0, \pi.$}
\label{bound}
\ee
On the horizon Eqns. (\ref{m}) and (\ref{mm}) become parabolic and have
the forms
\ben \label{h1}
{1 \over 2E} \p_{r} X \mid_{r = 2E} &=&
-{1 \over  2E \left ( 2E - {q^2 \over E} \right ) \sin \theta}
\p_{\theta}[\sin \theta \p_{\theta} X] + {N P^2 X \over
2E \left ( 2E - {q^2 \over E} \right ) \sin^2 \theta}
+ {1 \over 2} X(X^2 - 1), \\
{1 \over 2E} \p_{r} P \mid_{r = 2E} &=&
- {\sin \theta \over 2E \left ( 2E - {q^2 \over E} \right )}
\p_{\theta}[ csc \theta \p_{\theta} P ] + {X^2 P \over \beta}.
\label{h2}
\een
In order to solve
numerically Eqs. (\ref{m}) and (\ref{mm}) we use the
simultaneous over-relaxation method \cite{pftv} modified in the way
described in \cite{gg} to handle boundary conditions (\ref{h1}) and
(\ref{h2}) on the horizon. Eqns. are solved on a uniformly spaced
polar grid ${(r_i,\theta_i)}$, with boundaries at $r_{in}=2E$, outer
radius $r_{out} \gg 2E$ (we usually choose $r_{out}=10E$), and $\theta$
ranging from $0$ to $\pi$. The mesh is divided into $200 \times 200$ cells
(only for extreme black holes we use $300 \times 300$ cells).
Following the notation from Ref.\cite{gg}, the finite difference scheme for
Eqs. (\ref{m}) and (\ref{mm}) has the form
\ben \label{diff1}
X_{00}={\left ( 1-{2 E \over r} \right ) 
{X_{+0}+X_{-0} \over (\Delta
r)^2} + {1 \over r (r - {q^2 \over E})}
{X_{0+}+X_{0-} \over (\Delta
\theta)^2} + \left [ {1 \over r} 
+ {r -2E \over r (r - {q^2 \over
E})} \right ] {X_{+0}-X_{-0} \over 2 \Delta r} + 
{\cot \theta \over r (r - {q^2
\over E})}{X_{0+}-X_{0-} 
\over 2 \Delta \theta} \over \left ( 1 - {2E
\over r} \right ) {2 \over (\Delta r)^2} 
+ {2 \over r (r - {q^2 \over
E}) (\Delta \theta)^2} 
+ {1 \over r (r - {q^2 \over E})}{\left ( {P_{00}
N \over \sin \theta} \right )}^2 
+ {1 \over 2} (X_{00}^2 - 1)} \\
P_{00}={\left ( 1-{2 E \over r} 
\right ) {P_{+0}+P_{-0} \over (\Delta
r)^2} + {1 \over r (r - {q^2 \over E})}
{P_{0+}+P_{0-} \over (\Delta
\theta)^2} + {2E \over r} 
{P_{+0}-P_{-0} \over 2 \Delta r} - {\cot
\theta \over r (r - {q^2
\over E})}{P_{0+}-P_{0-} 
\over 2 \Delta \theta} \over \left ( 1 - {2E
\over r} \right ) {2 \over (\Delta r)^2} 
+ {2 \over r (r - {q^2 \over
E}) (\Delta \theta)^2} + {1 \over \beta} X_{00}^2},
\label{diff2}
\een
while on the horizon we have
\ben \label{bound1}
X_{00}={{1 \over \Delta r} X_{+0} 
+ {\cot \theta \over (2E - {q^2
\over E})}{X_{0+}-X_{0-} 
\over 2 \Delta \theta} + {1 \over (2E - {q^2
\over E})}{X_{0+}
+X_{0-} \over (\Delta \theta)^2} \over {1 \over
\Delta r} + E(X_{00}^2 - 1) 
+ {1 \over  (2E - {q^2 \over E})}{\left ( {P_{00}
N \over \sin \theta} \right )}^2 
+ {2 \over  (2E - {q^2 \over E})
(\Delta \theta)^2}}\\
P_{00}={{1 \over \Delta r} P_{+0} 
- {\cot \theta \over (2E - {q^2
\over E})}{P_{0+}-P_{0-} 
\over 2 \Delta \theta} + {1 \over (2E - {q^2
\over E})}{P_{0+}+P_{0-} \over (\Delta \theta)^2} 
\over {1 \over
\Delta r} + {1 \over \beta} 2 E X_{00}^2 
+ {2 \over  (2E - {q^2 \over E})
(\Delta \theta)^2}}
\label{bound2}
\een
Note that for $q=0$ we reproduce Eqns. obtained by Achucarro {\it
et al.} in Ref.\cite{gg} and we 
use their results to check our numerical code
obtaining the excellent agreement.
In order to begin
the numerical calculations, we first set the boundary conditions
according 
to (\ref{bound}). On the horizon, we initially set $X=1$ and
$P=0$. Then, the grid is over relaxed, i.e., values of $P$ and $X$ at each
grid point are updated with new values $X_{00}=\omega X_{new} +
(1-\omega) X_{00}$. The over relaxation parameter $\omega$, $( 1< \omega <
2)$ is found by trying some values and choosing the optimal one (we don't
use Chebyshev acceleration in our calculations). After an each iteration,
values of the fields on the horizon are updated according to
(\ref{bound1}) and (\ref{bound2}). The whole step is repeated until
reaching the convergence.
\par
The examples of our numerical investigations are depicted in
Figs.1-14. We also performed panels which enable to see cuts around
the horizon and along the equator for the $X$ and $P$
fields in the electrically charged dilaton black hole background. These
results confirm our analytical considerations given above.
\par
Figs.1-6 presented calculations for the winding number $N = 1$ and
different values of $E$ and $q$ parameters. In Figs.1-2 the Bogomolnyi
parameter $\beta$ is set equal to unity, i.e.,
the magnetic and the Higgs flux tubes are of the same size.
Figs.7-8 are performed for the increasing winding number, we set
$N=100$. The string is still noticeably pinched.
For all
these parameters the $X$ and $P$ fields are passing right through
the electrically charged dilaton black hole.
\par
The next problem, will be to analyze the behavior of the $X$ and $P$
fields in the background of the extreme black hole. 
In Sec.II, we obtained the analytical results, assuming
that the vortex size is large compared to
the black hole size.
Now, we shall deal with
the problem using the numerical code. Figs.9-14 show our sample results
and confirm the previous considerations. To begin with, we took into
considerations the extreme black holes with $E = 1, 10.0, 20.0$ and
the adequate values of $q$. 
In Figs.13-14 we take $\beta=1$, as in the previous non-extreme case.
In all cases, the $X$ field wraps around
the black hole horizon.
As far as the $P$ field is concerned, one
has the same situation. The $P$ fields wrap around the black hole horizon
and there is no penetration at all.
Having in mind the relation between $P$ and $\tB_{\theta \phi}$, it is
obvious that no magnetic flux is crossing the horizon.
The extreme electrically charged
dilaton black hole behaves like a perfect diamagnet. The same
situation was revealed in the case of the extreme Reissner-Nordstr\"om
black hole, in Ref.\cite{cha}. 
\par
One can conclude that,
studying the behavior of a non-extreme
electrically charged dilaton black hole and a black hole which is
quite close to the extremal case, we have the situation that the $X$
and $P$ fields flow through the outer horizon of the black hole.
On the other hand, in the case of the extreme electrically charged
dilaton black hole there is
an expulsion of the fields of the vortex from the horizon of the black hole.
\section{Gravitating strings}
In this Sec. our main task will be to find the metric of an Abelian
Higgs vortex passing through an electrically charged dilaton black
hole. To deal with the problem, we shall use the iterative procedure
and assume {\it the thin string} limit, i.e., $E \gg 1$.
\par
As in Ref.\cite{gg}, to consider the gravitational effect of the
Abelian Higgs vortex sitting on the electrically charged black hole,
one needs to take into account a general axially symmetric line element
of the form
\be
ds^2 = - e^{2 \psi} dt^2 + \tal^2 e^{2 \psi} d\phi^2 + e^{- 2\psi + 2
\ga} (d\tr^2  + dz^2).
\ee
We introduce the rescaled coordinates \cite{gg}
\ben
\rho &=& \sqrt{\la} \eta \tr,\\
\zeta &=& \sqrt{\la} \eta z,\\
\alpha &=& \sqrt{\la} \eta \tal .
\label{co}
\een
In terms of them the Einstein Eqns. yield
\be
\alpha_{, \zeta  \zeta} + \alpha_{,\rho \rho} = \ep \sqrt{- g}
\left ( \hT_{\zeta}{}{}^{\zeta} + \hT_{\rho}{}{}^{\rho} \right ),
\label{m1}
\ee
\be
( \alpha \psi_{,\zeta} )_{,\zeta} + ( \alpha \psi_{,\rho} )_{,\rho}
=  - {1 \over 2} \ep \sqrt{- g}
\left ( \hT_{0}{}{}^{0} -  \hT_{\zeta}{}{}^{\zeta} -  
\hT_{\rho}{}{}^{\rho} -  \hT_{\phi}{}{}^{\phi}
 \right ),
\label{m2}
\ee
\be
- \ga_{,\rho} (\alpha^2_{, \rho} + \alpha^2_{, \zeta}) +
\alpha \alpha_{, \rho} (\psi^2_{, \rho} - \psi^{2}_{, \zeta})  +
2 \alpha \alpha_{, \zeta} \psi_{, \rho} \psi_{,\zeta}
+ \alpha_{, \rho} \alpha_{,\rho \rho} + \alpha_{,\zeta} \alpha_{,\rho
\zeta} = \ep \sqrt{- g}
\left ( \alpha_{,\rho} \hT_{\zeta}{}{}^{\zeta} - 
\alpha_{, \zeta} \hT_{\zeta}{}{}^{\rho} \right ),
\label{m3}
\ee
\be
\ga_{,\zeta} (\alpha^2_{, \rho} + \alpha^2_{, \zeta}) -
\alpha \alpha_{, \zeta} (\psi^2_{, \rho} - \psi^{2}_{, \zeta})  -
2 \alpha \alpha_{, \rho} \psi_{, \rho} \psi_{,\zeta}
+ \alpha_{, \zeta} \alpha_{,\rho \rho} - \alpha_{,\rho} \alpha_{,\rho
\zeta} = \ep \sqrt{- g}
\left ( \alpha_{,\zeta} \hT_{\zeta}{}{}^{\zeta} - 
\alpha_{, \rho} \hT_{\zeta}{}{}^{\rho} \right ),
\label{m4}
\ee
\be
\ga_{,\zeta \zeta} + \ga_{, \rho \rho} + \psi^2_{, \zeta} +
\psi^2_{,\rho} = \ep \sqrt{- g}
\left ( e^{2 \ga - 2 \psi} \hT_{\phi}{}{}^{\phi} \right ),
\label{m5}
\ee
where $\ep = 8\pi G\eta^2$ which is assumed to be small. This
assumption is well justified because, e.g., for the
GUT string
$\ep \leq 10^{-6}$. The rescaled energy momentum tensor,
i.e., $\hT_{\alpha}{}{}^{\beta} = {T_{\alpha}{}{}^{\beta} \over \la
\eta^4}$ is as follows:
\ben
\hT_{0}{}{}^{0} &=& - e^{-2(\ga - \psi)} \left (
X^2_{,\rho} + X^2 _{,\zeta} \right ) - {X^2 P^2 e^{2\psi} \over
\alpha^2} - {\beta \over \alpha^2} e^{-2\ga + 4\psi}
\left ( P^2_{, \rho} + P^2_{, \zeta} \right ) - V(X), \\
\hT_{\phi}{}{}^{\phi} &=& - e^{-2(\ga - \psi)} \left (
X^2_{,\rho} + X^2 _{,\zeta} \right ) + {X^2 P^2 e^{2\psi} \over
\alpha^2} + {\beta \over \alpha^2} e^{-2\ga + 4\psi}
\left ( P^2_{, \rho} + P^2_{, \zeta} \right ) - V(X), \\
\hT_{\rho}{}{}^{\rho} &=& e^{-2(\ga - \psi)} \left (
X^2_{,\rho} - X^2 _{,\zeta} \right ) - {X^2 P^2 e^{2\psi} \over
\alpha^2} + {\beta \over \alpha^2} e^{-2\ga + 4\psi}
\left ( P^2_{, \rho} - P^2_{, \zeta} \right ) - V(X), \\
\hT_{\zeta}{}{}^{\zeta} &=& - e^{-2(\ga - \psi)} \left (
X^2_{,\zeta} - X^2 _{,\rho} \right ) - {X^2 P^2 e^{2\psi} \over
\alpha^2} + {\beta \over \alpha^2} e^{-2\ga + 4\psi}
\left ( P^2_{, \zeta} - P^2_{,\rho} \right ) - V(X), \\
\hT_{\rho}{}{}^{\zeta} &=& 2 e^{-2(\ga - \psi)} \left (
X_{,\rho}X_{,\zeta} + {\beta \over \alpha^2}
P_{, \rho}P_{, \zeta} \right ), 
\een
where $V(X) = {(X^2 -1)^2 \over 4}$.
\par In order to solve the Einstein Eqns. we shall use the
iterative procedure expanding Eqns. of motion in terms of the
quantity $\ep$. Our starting point is to rewrite the line element
of a charged dilaton black hole into $\tr$ and $z$ coordinates
defined as
\ben
\tr^2 &=& (\tir - 2GM) \left ( \tir - {Q^2  \over GM} \right )
\sin^2 \theta, \\
z &=& \left ( \tir - GM - {Q^2 \over 2GM} \right ) \cos \theta.
\een
In the above coordinates the metric of the electrically charged
dilaton black hole takes the form
\be
ds^2 = - e^{2 \psi_{0}} dt^2 + \tr^2 e^{- 2 \psi_{0}} d\phi^2 + e^{-
2\psi_{0}  + 2
\gamma_{0}} (d \tr^2  + dz^2),
\ee
where
\ben
e^{- 2 \psi_{0}} &=& {\tir \over \tir - 2GM},\\
e^{- 2 \psi_{0} + 2 \gamma_{0}} &=& 
{\left ( \tir - {Q^2 \over GM} \right ) \over
A} ,\\
\tr^2 &=& {\tilde \alpha}_{0}^2, \\
A &=& (\tir - 2GM) \left ( \tir - {Q^2  \over GM} \right )
\cos^2 \theta + {\sin^2 \theta \over 4} \left [
(\tir - 2GM) +  \left ( \tir - {Q^2  \over GM} \right ) \right ]^2.
\een
To the zeroth order, they will constitute our background solution.
\par
We proceed to check if the energy momentum tensor admits a geodesic
shear free event horizon. Taking into account the relation
$R_{\alpha \beta} l^{\alpha}l^{\beta} = 0$ \cite{car}, we
reach to the conclusion
that it is equivalent to the condition $T_{0}{}{}^{0} - T_{r}{}{}^{r} =
0$. This relation in the rescaled coordinates (\ref{co}) has the form
\be
\hT_{0}{}{}^{0} - \hT_{\rho}{}{}^{\rho} = 0.
\ee
Using the coordinates
\be 
R = \sqrt{r (r - {q^2 \over E})} \sin \theta = \rho e^{-\psi_{0}}
\ee
and the exact form of the energy momentum tensor, we finally get
\be
\hT_{0}{}{}^{0} - \hT_{\rho}{}{}^{\rho} =
- 2 e^{-2(\ga - \psi)} \left ( {dR \over d \rho} \right )^2
\left [ {\beta \over R^2} \left ( \dR P \right )^2 + 
\left ( \dR X \right )^2 \right ].
\label{hor}
\ee
In Eqn. (\ref{hor}) all terms remain finite and non-zero as $\rho
\rightarrow 0$, except $R_{,\rho}$. Using the exact form of
$R_{, \rho}$ and $R_{,\zeta}$ derivatives, namely
\be
R_{,\rho} = {\p r \over \p \rho} {\p R \over \p r} + {\p \theta \over
\p \rho}{\p R \over \p \theta} =
{\rho \over R} \left [ {2r - {q^2 \over E} \over (r - 2E) + 
(r - {q^2 \over E})} + {r \over r - 2E} \right ],
\ee
and
\be
R_{,\zeta} = {\p r \over \p \zeta} {\p R \over \p r} + {\p \theta \over
\p \zeta}{\p R \over \p \theta} =
{(2r - {q^2 \over E}) \sin \theta \over 2\sqrt{r(r - {q^2 \over E})}
\cos \theta} - {\sqrt{r(r - {q^2 \over E})} \cos \theta \over
{\sin \theta \over 2} (2r - 2E -{q^2 \over E})}.
\ee
One can see that,
on the horizon where $\rho \rightarrow 0$, ${dR \over d \rho}
\rightarrow 0$ 
which implies that
$\hT_{0}{}{}^{0} - \hT_{\rho}{}{}^{\rho} = 0.$
Then, we draw the conclusion that, at least in a linearized method
of considering the problem of an Abelian Higgs vortex on an
electrically charged dilaton black hole, there is no gravitational
obstacles of painting the vortex on the horizon of the black hole under
consideration.
\vspace{0.5cm}
Near the core of the string where $\sin \theta \approx \cO(E^{-1})$,
we have the following relation:
\be
R_{,\rho}^{2} + R_{,\zeta}^{2} \sim e^{2\ga_{0} - 2\psi_{0}}.
\label{rr}
\ee 
This relation implies that near the core
of the string, to the zeroth order, the energy momentum tensor  reads
as follows:
\ben
\hT_{(0) 0}^{}{}{0} &=& - V(X_{0}) - \left ( \dR X_{0} \right )^{2}
- {X_{0}^2 P_{0}^2 \over R^2} - {\beta \over R^2} \left ( \dR P_{0} \right
)^2 , \\
\hT_{(0) \phi}^{}{}{\phi} &=& - V(X_{0}) - \left ( \dR X_{0} \right )^{2}
+ {X_{0}^2 P_{0}^2 \over R^2} + {\beta \over R^2} \left ( \dR P_{0} \right
)^2 , \\
\hT_{(0) \rho}^{}{}{\rho} &=& - V(X_{0}) - \left ( \dR P_{0} \right )^{2}
+ e^{-2(\ga_{0} - \psi_{0})} \left [
{\beta \over R^2} \left ( \dR P_{0} \right )^2 + \left (
\dR X_{0} \right )^2 \right ] (R_{,\rho}^2 - R_{,\zeta}^2), \\
\hT_{(0) \zeta}^{}{}{\zeta} &=& - V(X_{0}) - \left ( \dR P_{0} \right )^{2}
- e^{-2(\ga_{0} - \psi_{0})} \left [
{\beta \over R^2} \left ( \dR P_{0} \right )^2 + \left (
\dR X_{0} \right )^2 \right ] (R_{,\rho}^2 - R_{,\zeta}^2), \\
\hT_{(0) \rho}^{}{}{\zeta} &=&
2 e^{-2(\ga_{0} - \psi_{0})} \left [
{\beta \over R^2} \left ( \dR P_{0} \right )^2 + \left (
\dR X_{0} \right )^2 \right ] R_{,\rho}R_{,\zeta}.
\een
which is the purely function of $R$. As in 
the Schwarzschild case \cite{gg}, this
strongly suggests to look for the metric perturbations as a function
of $R$.
\par
Writing Eqns. of motion to the first order of $\ep$, one obtains
\be
\alpha_{1, \rho \rho} + \alpha_{1, \zeta \zeta} = -
\alpha_{0} e^{2\ga_{0} - 2\psi_{0}}
\left ( 2V(X_{0}) + {2 X_{0}^2 P_{0}^2 \over R^2} \right ),
\label{mm1}
\ee
\be
\alpha_{1, \rho} \psi_{0, \rho} + (\rho \psi_{1, \rho})_{, \rho} +
\alpha_{1, \zeta} \psi_{0, \zeta} + \rho \psi_{1, \zeta \zeta}  =
- {1 \over 2} \alpha_{0} e^{2\ga_{0} - 2\psi_{0}}
\left ( 2V(X_{0}) - {2 \beta \over R^2} (P_{0}')^2 \right ),
\label{mm2}
\ee
\be
\ga_{1, \zeta \zeta} + \ga_{1, \rho \rho} + 2\psi_{0, \zeta}\psi_{1,
\rho} = e^{2\ga_{0} - 2\psi_{0}}
\left ( - (X_{0}')^2 + {X_{0}^2 P_{0}^2 \over R^2}  +
{\beta \over R^2} (P_{0}')^2 - V(X_{0}) \right ).
\label{mm3}
\ee
Taking into account (\ref{rr}), one has that \cite{gg},
$\p_{\zeta}^{2} + \p_{\rho}^{2} = e^{2 (\ga_{0} - \psi_{0})}
\left [ {d^2 \over dR^2} + \cO ( {1\over E^2} ) \right ]$.
The close inspection of Eqn. (\ref{mm1}) reveals that, we can write
$\alpha_{1}$ as a function of $\rho$ and $R$, namely
$\alpha_{1} = \rho a(R)$. $a(R)$ yields
\be
{d^2 \over dr^2}a(R) + {2\over R} \dR a = \hT_{\rho}{}{}^{\rho}
+ \hT_{\zeta}{}{}^{\zeta}.
\ee
Then, one reaches to the following expression for $a(R)$:
\be
a(R) = \int_{R}^{\infty} {1 \over R^2} dR \int_{0}^{R} R'^2
\left ( - 2V(X_{0}) - {2X_{0}^2 P_{0}^2 \over R'^2} \right ) dR'.
\label{a}
\ee
Eqn. (\ref{a}) can be rewritten as
\be
a(R) \sim - {A \over \ep} + {B \over \ep R},
\ee
where
\ben
A &=& \ep \int_{0}^{R} R (\hT_{\rho (0)}{}{}^{\rho} +
\hT_{\zeta (0)}{}{}^{\zeta}) dR, \\
B &=& \ep \int_{0}^{R} R (\hT_{\rho (0)}{}{}^{\rho} +
\hT_{\zeta (0)}{}{}^{\zeta}) dR.
\een
Using the form of $\alpha_{1}$ and setting $\psi_{1} = \psi_{1}(R)$,
we have
\be
\psi_{1}(R) = - {1 \over 2}
\int_{R}^{\infty} {1 \over R} dR \int_{0}^{R} R'
\left (  2V(X_{0}) - {2 \beta \over R'^2} \left ( {dP_{0} \over dR}
\right )^2 \right ) dR'.
\ee
While for $\ga_{1}(R)$, one arrives at the expression of the form
\be
\ga_{1}(R) = \int_{R}^{\infty} dR \int_{0}^{R}
\hT_{\phi}{}{}^{\phi} dR'= 2 \psi_{1}(R).
\ee
Having in mind the above corrections to the metric functions,
the asymptotic form of the metric yields
\be
 ds^2 \rightarrow e^C
\left [ -e^{2\psi_{0}}dt^2 + e^{2( \ga_{0} - 2 \psi_{0})}( dz^2 +
d\tr^2 ) \right ]
+ \tr^2 \left ( 1 - A + {B \over \sqrt{\la} \eta \tr e^{\psi_{0}}}
\right )^2 e^{- C} e^{-2\psi_{0}} d\phi^2,
\ee
or consistently with the coordinates (\ref{co}) is determined by
\ben
ds^2 \rightarrow e^{C} \left [ -
\left ( 1 - {2GM \over \tir} \right ) dt^2 +
{d \tir^2 \over \left ( 1 - {2GM \over \tir} \right ) } +
\tir \left ( \tir - {Q^2 \over GM} \right ) d \theta^2 \right ] 
+ \\ \nonumber
+ \tir \left ( \tir - {Q^2 \over GM} \right ) \left [
1 - A + {B \over \sqrt{\la} \eta \tir e^{\psi_{0}}}
\right ]^2 e^{-C} \sin^2 \theta d\phi^2,
\een
where $e^C = e^{2 \ep \psi_{1}}$. \\
The $B$-term represents the effect outside the range of the applicability of
the considered approximation \cite{gg}. Therefore, one should drop this term.
Rescaling the metric, $\hht = e^{C/2}t, \hhr = e^{C/2} \tir$ and
defining the quantities
\ben \label{mas}
\hhM = e^{C/2}M, \\
\hhQ = e^{C/2} Q, 
\label{lad}
\een
one gets the metric of the
electrically charged dilaton black hole with a string passing through
it, the metric of {\it the conical} electrically charged dilaton black
hole, namely
\be 
ds^2 =  -
\left ( 1 - {2G \hhM \over \hhr} \right ) d \hht^2 +
{d \hhr^2 \over \left ( 1 - {2G \hhM \over \hhr} \right ) } +
\hhr \left ( \hhr - {\hhQ^2 \over G\hhM} \right ) d \theta^2 +
\hhr \left (
\hhr -{\hhQ^2 \over G \hhM} \right ) (1- A)^2 e^{-2C} \sin^2 \theta
d\phi^2 .
\ee
As in the Schwarzschild case \cite{gg}, due to the presence of the radial
pressure term $e^{-C}$, we have the modified Schwarzschild mass
parameter at infinity (\ref{mas})  and the modified black hole charge
(\ref{lad}).
\par
The final issue which we wish to consider is a few remarks concerning
the thermodynamics of an electrically charged dilaton black hole with a
string passing through it. The temperature of a static black hole one
can obtain considering its behavior for imaginary value of time
\cite{te} 
\be
T = {1 \over 8 \pi G \hhM},
\ee
where the Boltzmann constant is set to unity.\\
The entropy may be inferred from the second law of thermodynamics, or
alternatively computed from the interpretation of the black hole as a
saddle-point contribution to the partition function. Then, it reads
\be
S = 2 \pi \hhM \left (
2 G \hhM - {\hhQ^2 \over G \hhM} \right ) (1 -A) e^{-C}.
\ee
The black hole temperature in unchanged comparing to the {\it
nonconical} case. Its value is in terms of the 
modified gravitational mass measured at infinity. On the other hand,
entropy of the black hole with a string is less than that of a black
hole of the same temperature without a string. It happened
because of the fact that the internal mass 
(which can be found by considering the black hole as formed by a
spherical shell of matter falling from infinity)
and gravitational one were no
longer equal. The very similar situation took place in the
Schwarzschild-vortex system \cite{gg}.

\section{Conclusions}
In our work we deal with the problem if the nontrivial topology of some
field configurations (such as strings, domain walls, textures) can act
as a hair for black holes. Namely, we considered an electrically
charged dilaton black hole originated in the low-energy string theory
and an Abelian Higgs vortex.
Assuming in our investigations a clear separation between the degrees
of freedom of the black hole-vortex configuration, we justify that
the effect of the vortex can be regarded from infinity as black hole
hair.
Our preliminary analytic studies in Sec.II, in the limit where the
vortex is thick compared to the horizon radius of the black hole,
reveal that the extreme electrically charged dilaton black hole
expells from its horizon all the
fields living in the core of a vortex. 
The extreme black hole behaves like a perfect diamagnet.
The similar kind of {\it a
Meissner effect} was revealed studying an extreme Reissner-Nordstr\"om-
vortex configuration \cite{cha}.
\par
By means of
the simultaneous over-relaxation method modified in order to handle the
boundary conditions,
we obtained the solutions of Eqns. of motion for an
Abelian Higgs vortex living in the background of the electrically
charged dilaton black hole. The Eqns. are elliptic outside the event
horizon, while they are parabolic on it. The numerical results confirm
the approximate calculations conducted in the previous section.
Summing it all up, we conclude that the fields of a vortex are always
expelled from the extreme black hole horizon. However, for the
non-extreme electrically charged dilaton black holes the fields $X$ and
$P$ are passing through the black hole horizon.
One should be aware that we do not take into account the back reaction
of the vortex on the geometry. This problem needs to be treated more
carefully and we hope to return to this question elsewhere.
\par
Starting with the background solution and the Nielsen-Olesen forms
of $X$ and $P$, by means of an iterative procedure of solving the field
Eqns., we found the metric of an electrically charged dilaton black hole
with a string passing through it.
The temperature of such a black hole is equal to the temperature of the
black hole without a string. It is measured in terms of the modified
Schwarzschild mass parameter measured at infinity. The entropy, in
turn, is less than the entropy of the black hole at the same
temperature without a string. The reason is caused by the fact that the
inertial mass is not equal to the gravitational one.

\vspace{3cm}
\noindent
{\bf Acknowledgements:}\\
We would like to thank Ruth Gregory for helpful remarks on various
occasions. R.M. acknowledges the support from the Foundation for Polish
Science Fellowship.


\pagebreak

\begin{figure}[p]
\begin{center}
\leavevmode
\epsfxsize=440pt
\epsfysize=440pt
\epsfbox{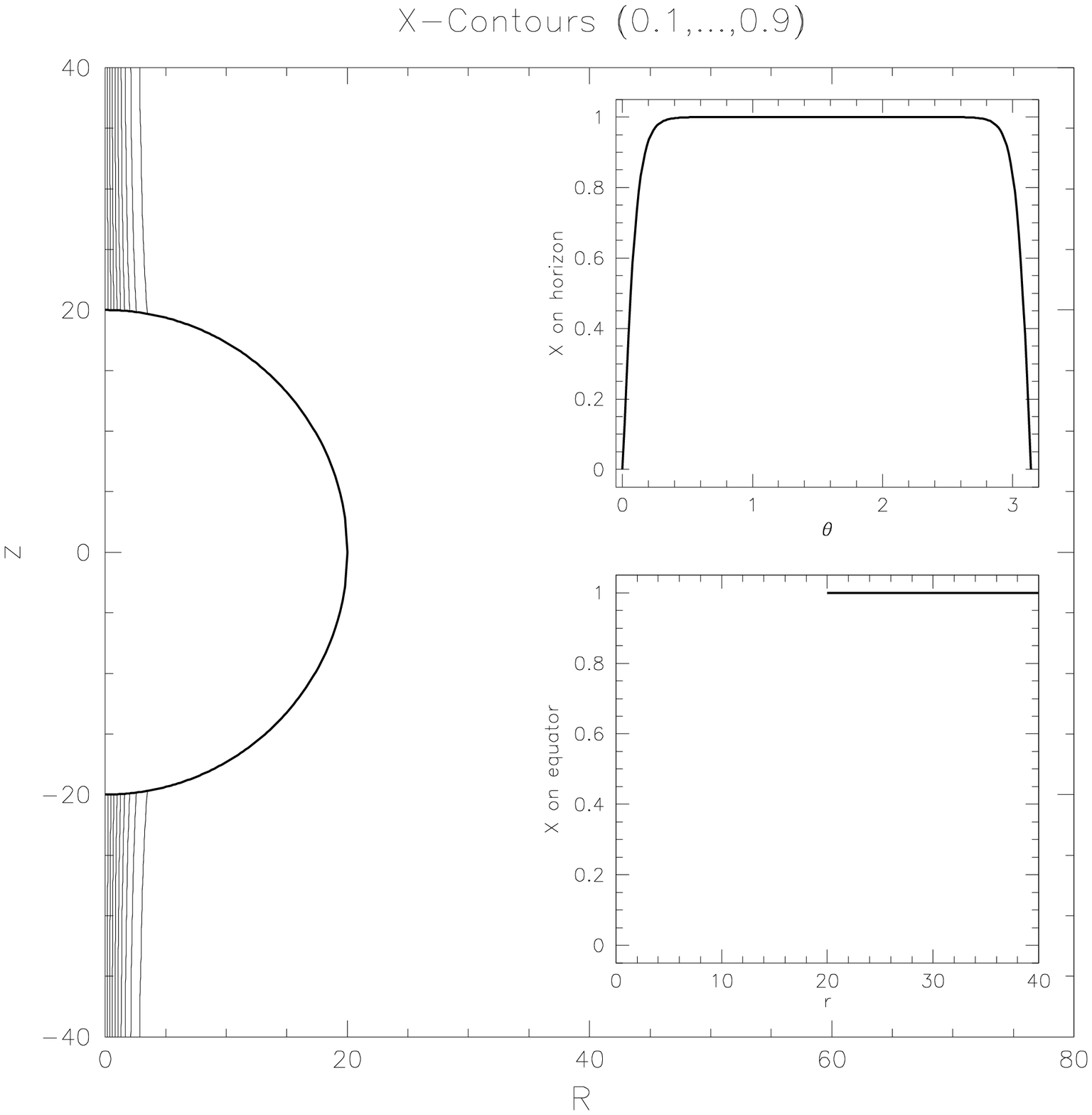}
\end{center}
\caption{\bf Numerical solution of the Nielsen-Olesen Eqns.
for the $X$ field with $N=1$, $\beta=1.0$, $E=q=10.0$.
The event horizon is indicated by a semicircle. Upper and lower panels
show cuts around the horizon and along the equator for the $X$ field in
the electrically charged dilaton black hole background.}
\label{fig1X}
\end{figure}
\begin{figure}
\begin{center}
\leavevmode
\epsfxsize=440pt
\epsfysize=440pt
\epsfbox{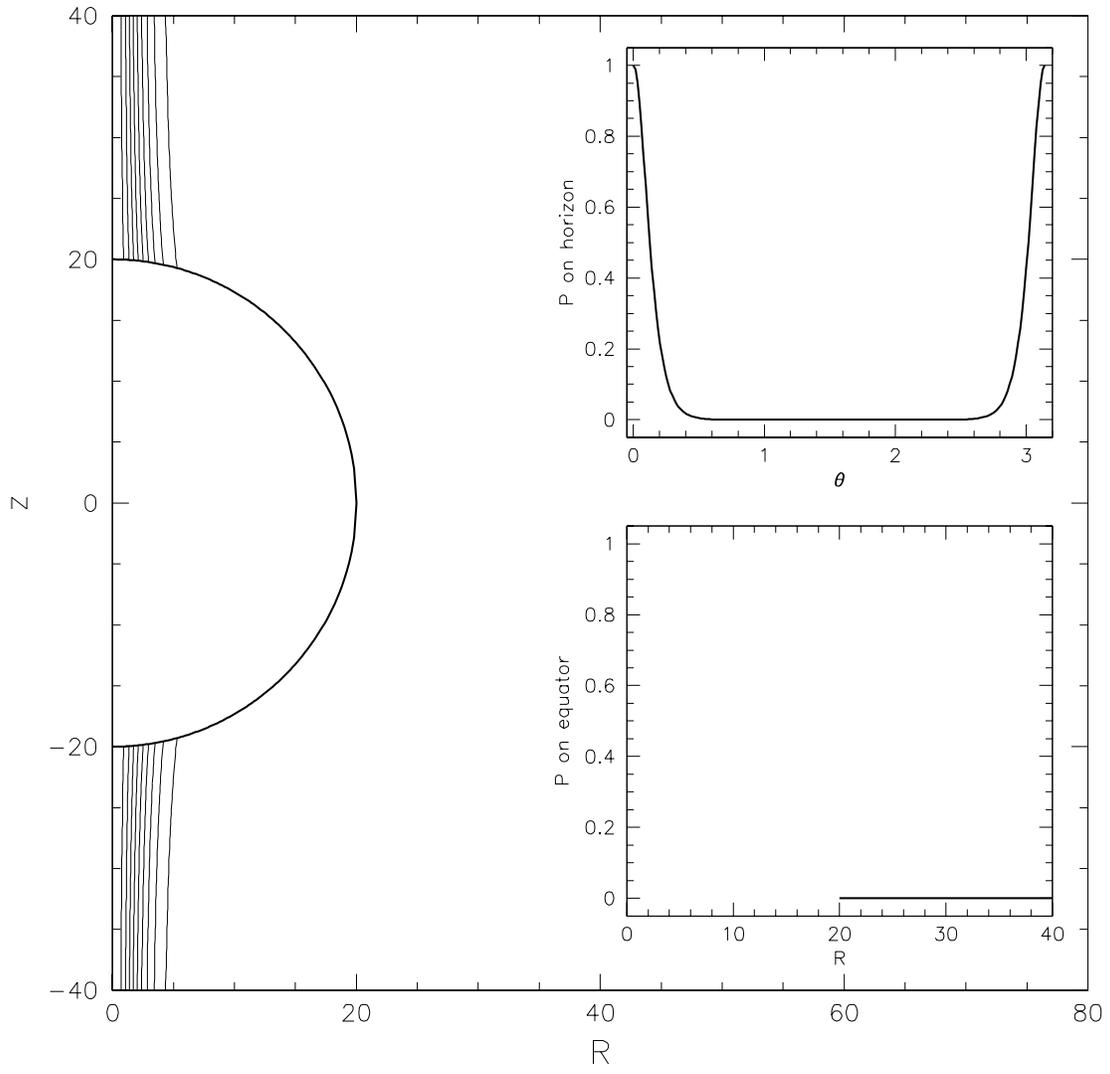}
\end{center}
\caption{\bf As in Fig.1 but for the $P$ field
with $N=1$, $\beta=1.0$, $E=q=10.0$.}
\label{fig1P}
\end{figure}
\begin{figure}
\begin{center}
\leavevmode
\epsfxsize=440pt
\epsfysize=440pt
\epsfbox{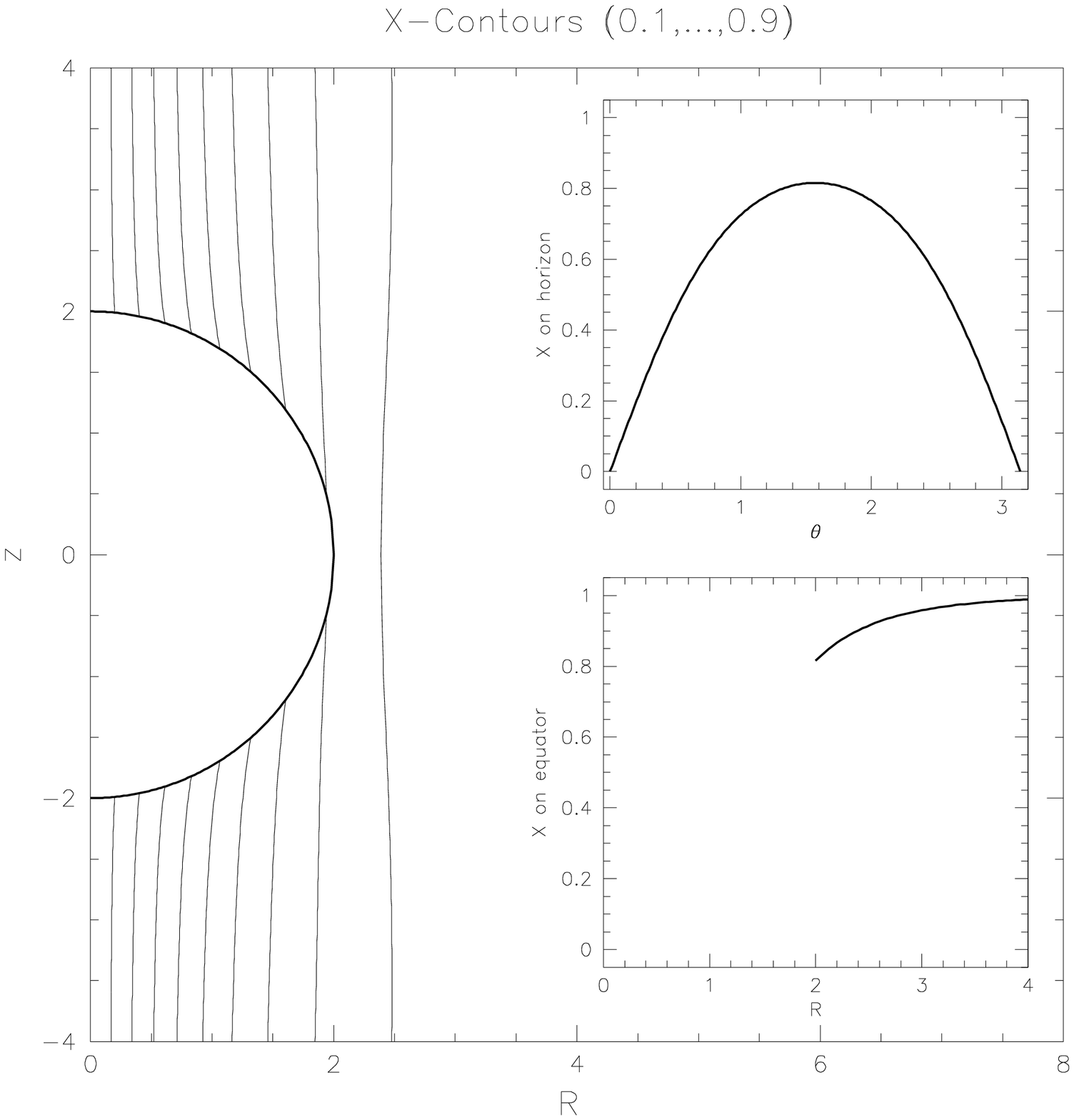}
\end{center}
\caption{\bf As in Fig.1 for the $X$ field
with $N=1$, $\beta=0.5$, $E=q=1.0$.}
\label{fig2X}
\end{figure}
\begin{figure}
\begin{center}
\leavevmode
\epsfxsize=440pt
\epsfysize=440pt
\epsfbox{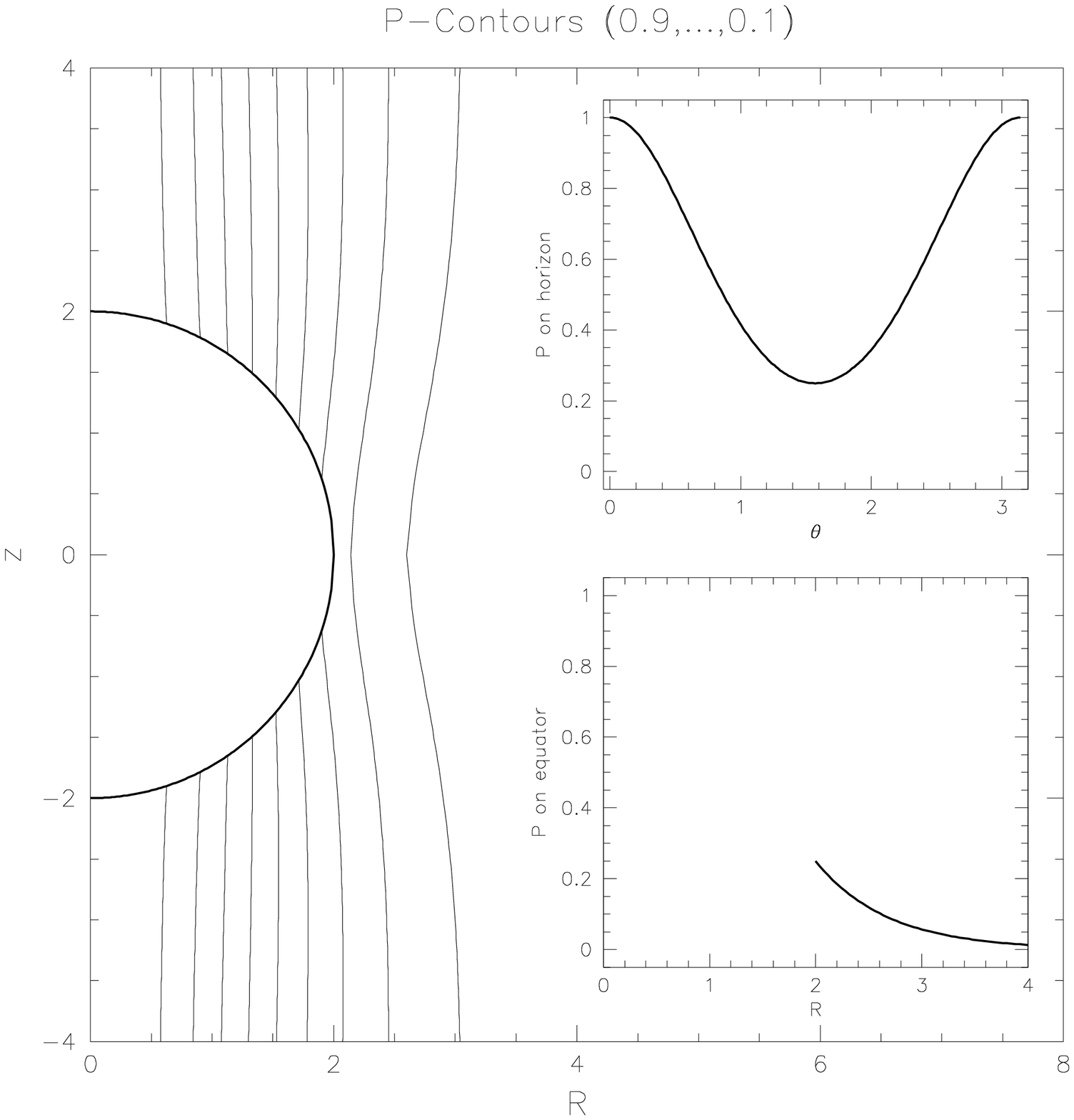}
\end{center}
\caption{\bf As in Fig.2 for the $P$ field
with $N=1$, $\beta=0.5$, $E=q=1.0$.}
\label{fig2P}
\end{figure}
\begin{figure}
\begin{center}
\leavevmode
\epsfxsize=440pt
\epsfysize=440pt
\epsfbox{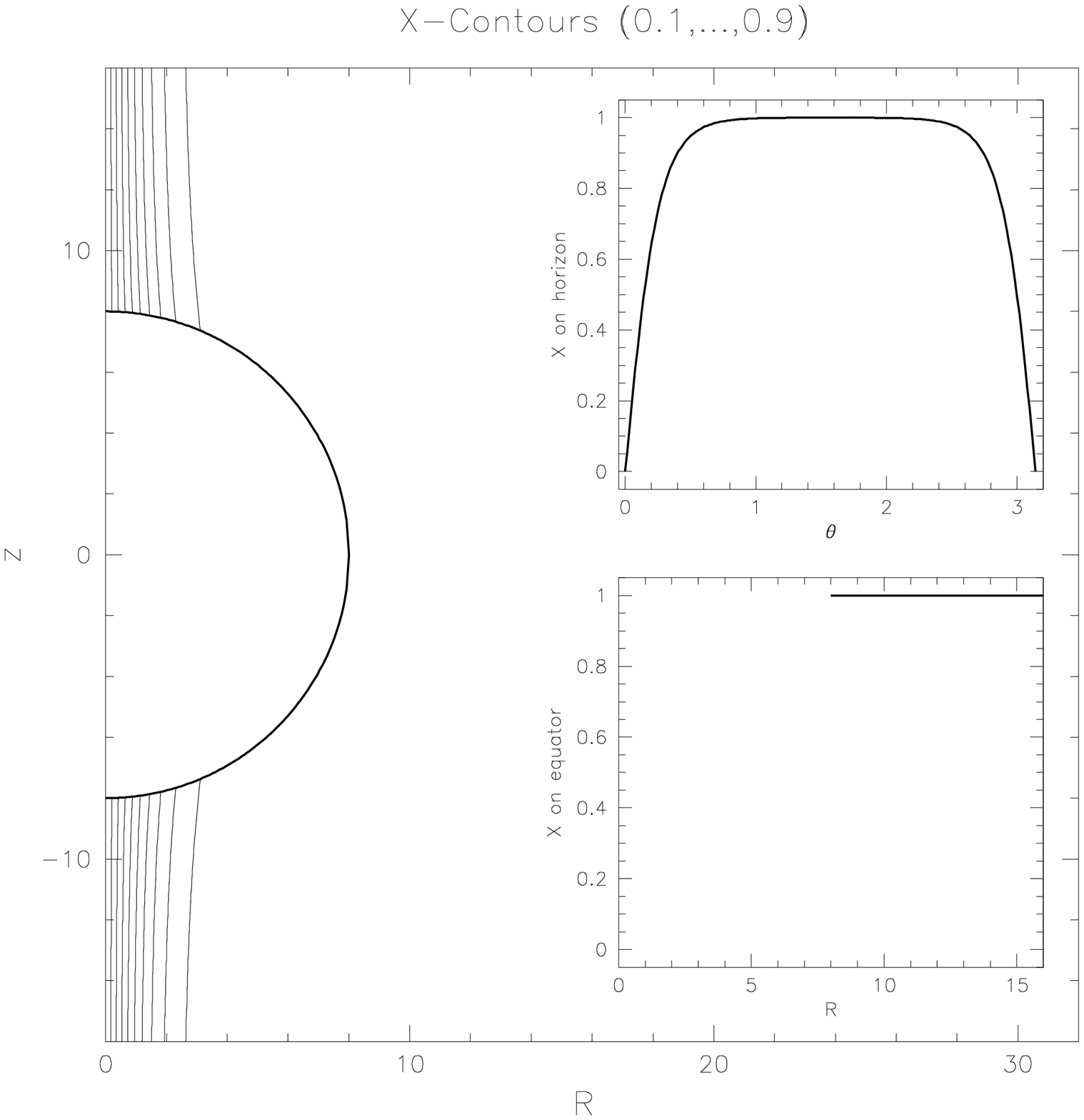}
\end{center}
\caption{\bf As in Fig.1 for the $X$ field
with $N=1$, $\beta=0.5$, $E=q=4.0$.}
\label{fig3X}
\end{figure}
\begin{figure}
\begin{center}
\leavevmode
\epsfxsize=440pt
\epsfysize=440pt
\epsfbox{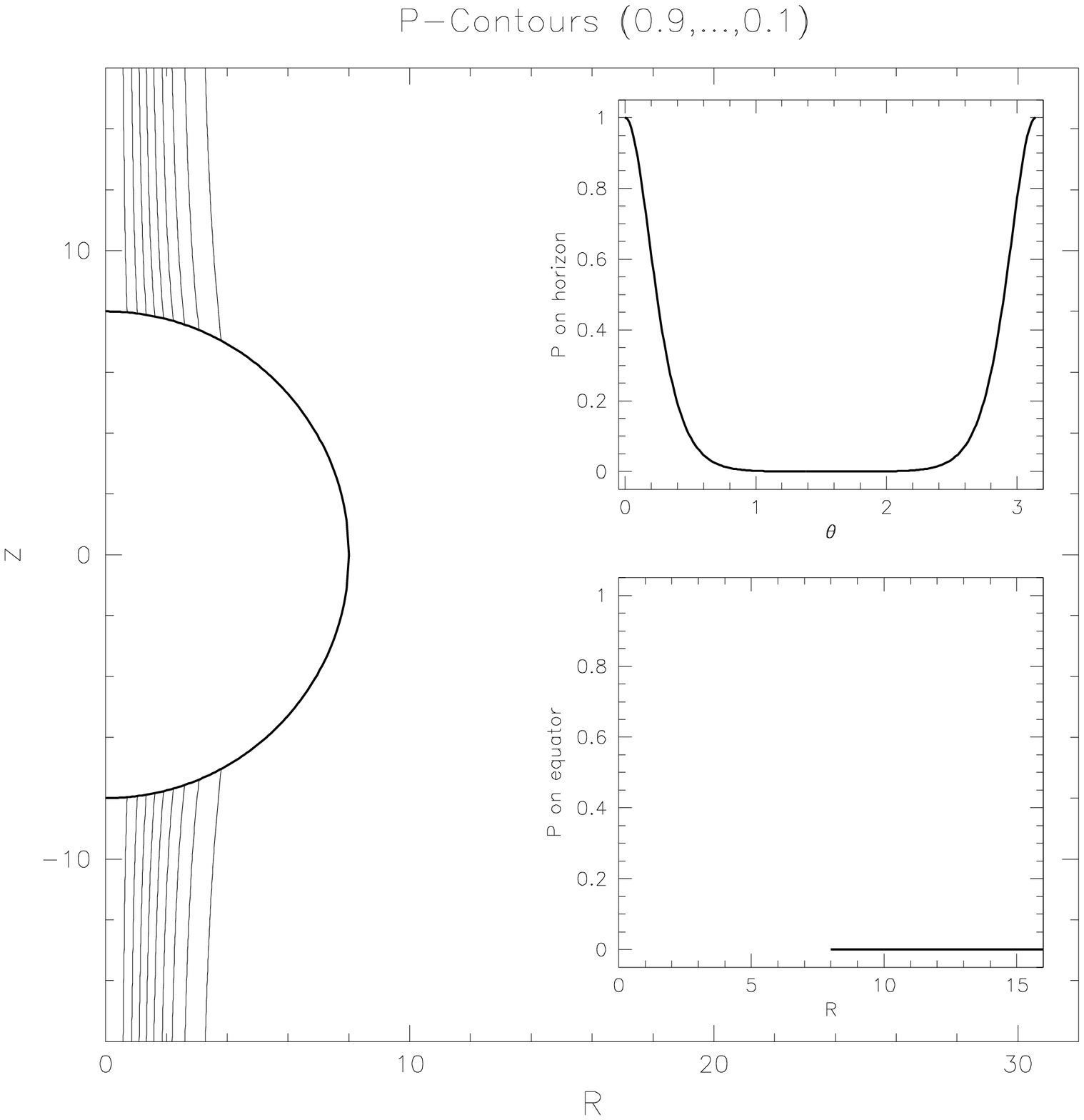}
\end{center}
\caption{\bf As in Fig.2 for the $P$ field
with $N=1$, $\beta=0.5$, $E=q=4.0$.}
\label{fig3P}
\end{figure}
\begin{figure}
\begin{center}
\leavevmode
\epsfxsize=440pt
\epsfysize=440pt
\epsfbox{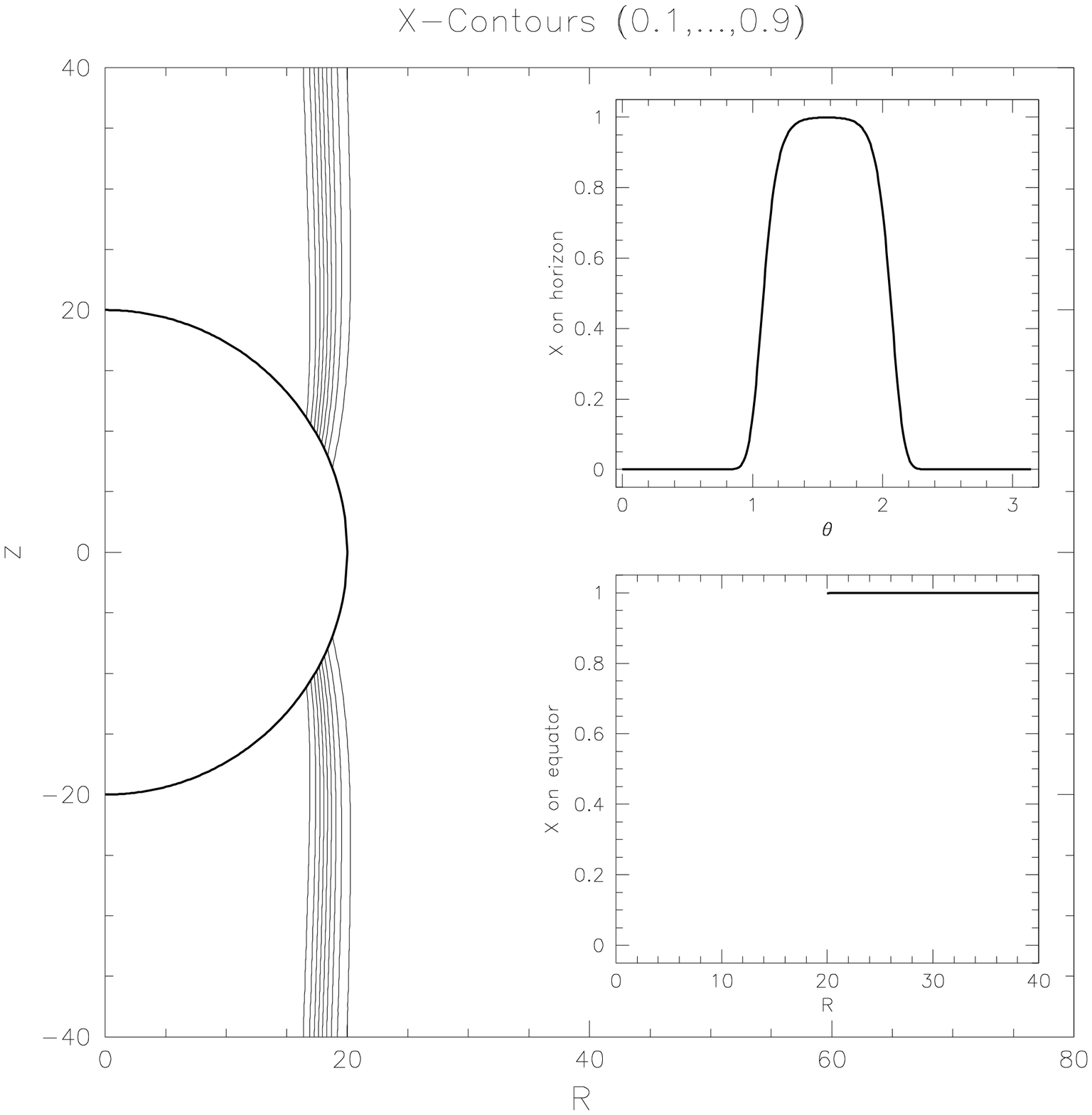}
\end{center}
\caption{\bf As in Fig.1 for the $X$ field
with $N=100$, $\beta=0.5$, $E=q=10.0$.}
\label{fig7X}
\end{figure}
\begin{figure}
\begin{center}
\leavevmode
\epsfxsize=440pt
\epsfysize=440pt
\epsfbox{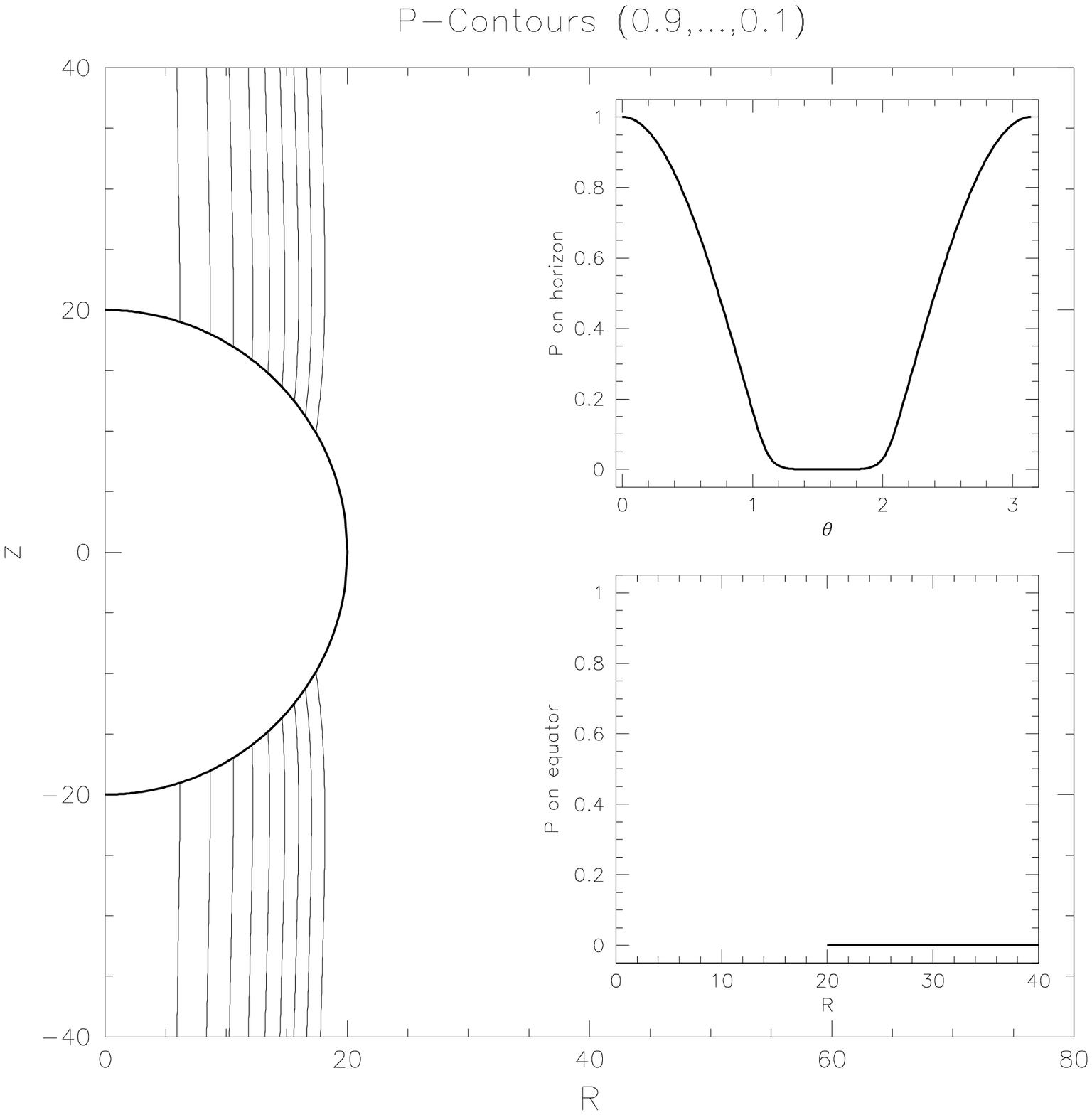}
\end{center}
\caption{\bf As in Fig.2 for the $P$ field
with $N=100$, $\beta=0.5$, $E=q=10.0$.}
\label{fig7P}
\end{figure}
\begin{figure}
\begin{center}
\leavevmode
\epsfxsize=440pt
\epsfysize=440pt
\epsfbox{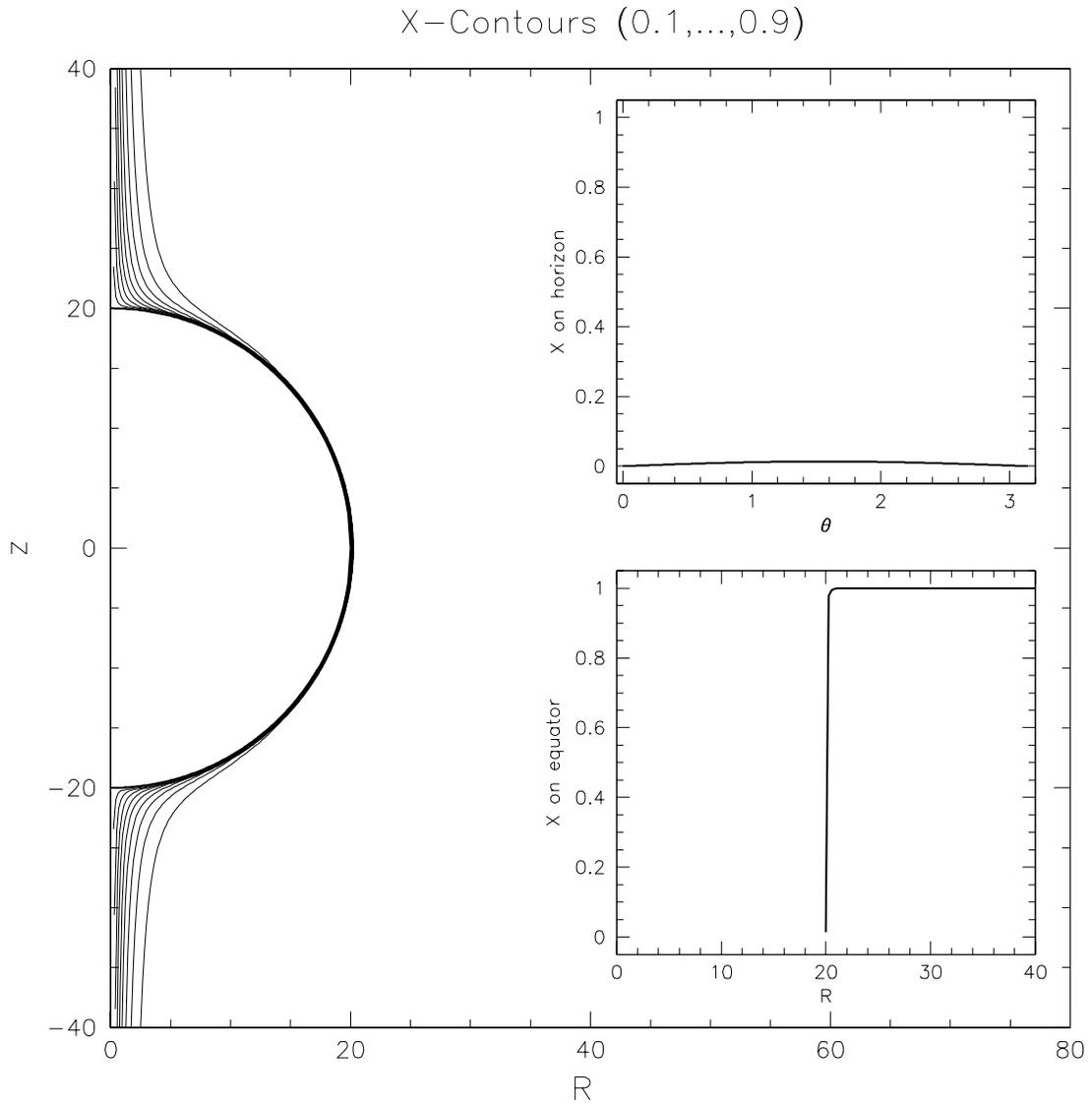}
\end{center}
\caption{\bf Numerical solution of the Nielsen-Olesen
Eqns. for the $X$ field
with $N=1$, $\beta=0.05$, $E=10.0$, $q=14.142 $ in background of the
extreme electrically charged dilaton black hole. One can 
notice the expulsion of the
$X$ field by the extreme black hole.}
\label{fig4X}
\end{figure}
\begin{figure}
\begin{center}
\leavevmode
\epsfxsize=440pt
\epsfysize=440pt
\epsfbox{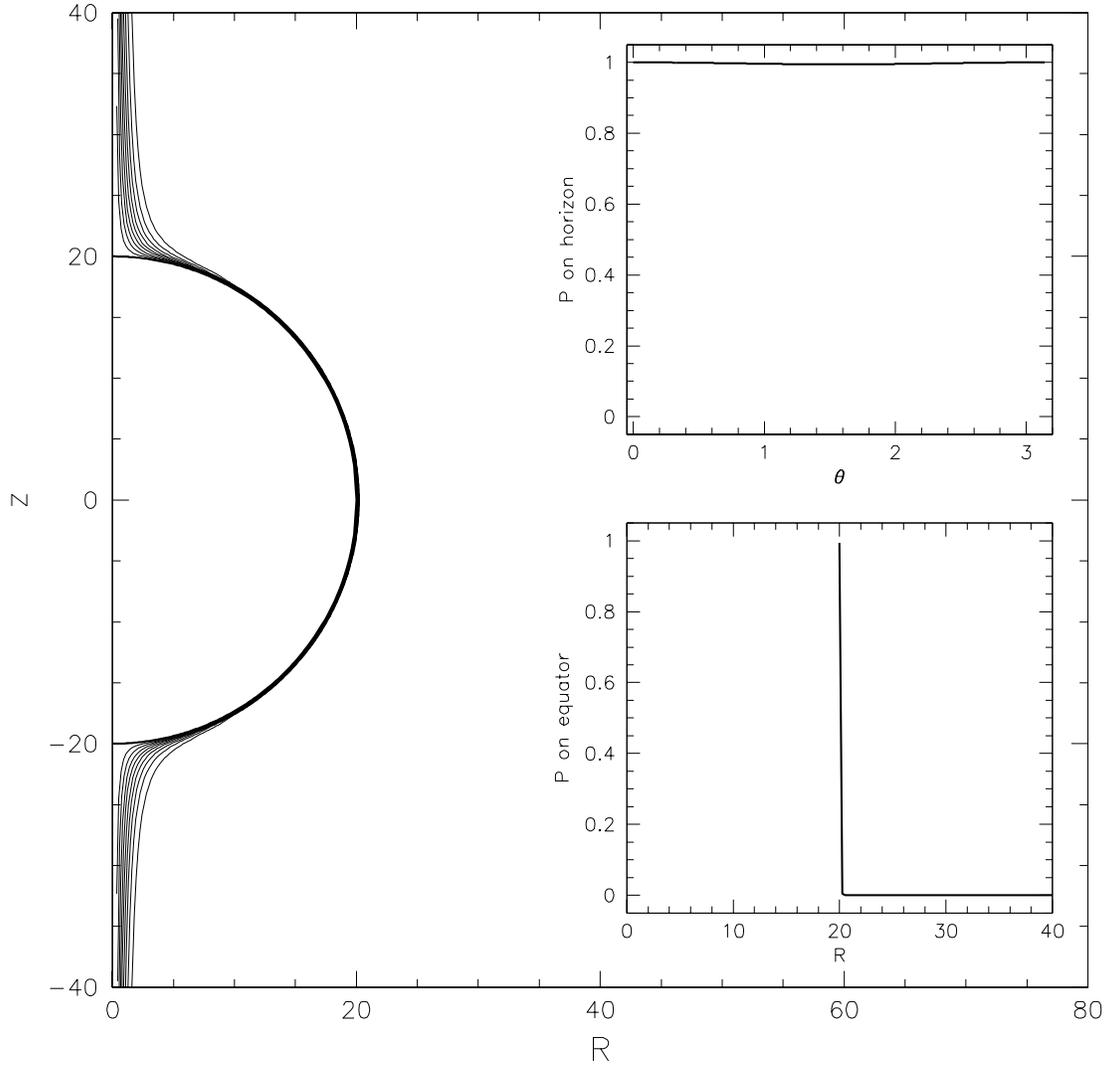}
\end{center}
\caption{\bf As in Fig.7 but for the $P$ field
with $N=1$, $\beta=0.05$, $E=10.0$, $q=14.142$.
The same situation, one has the expulsion of the $P$ field from the
extreme electrically charged dilaton black hole.}
\label{fig4P}
\end{figure}
\begin{figure}
\begin{center}
\leavevmode
\epsfxsize=440pt
\epsfysize=440pt
\epsfbox{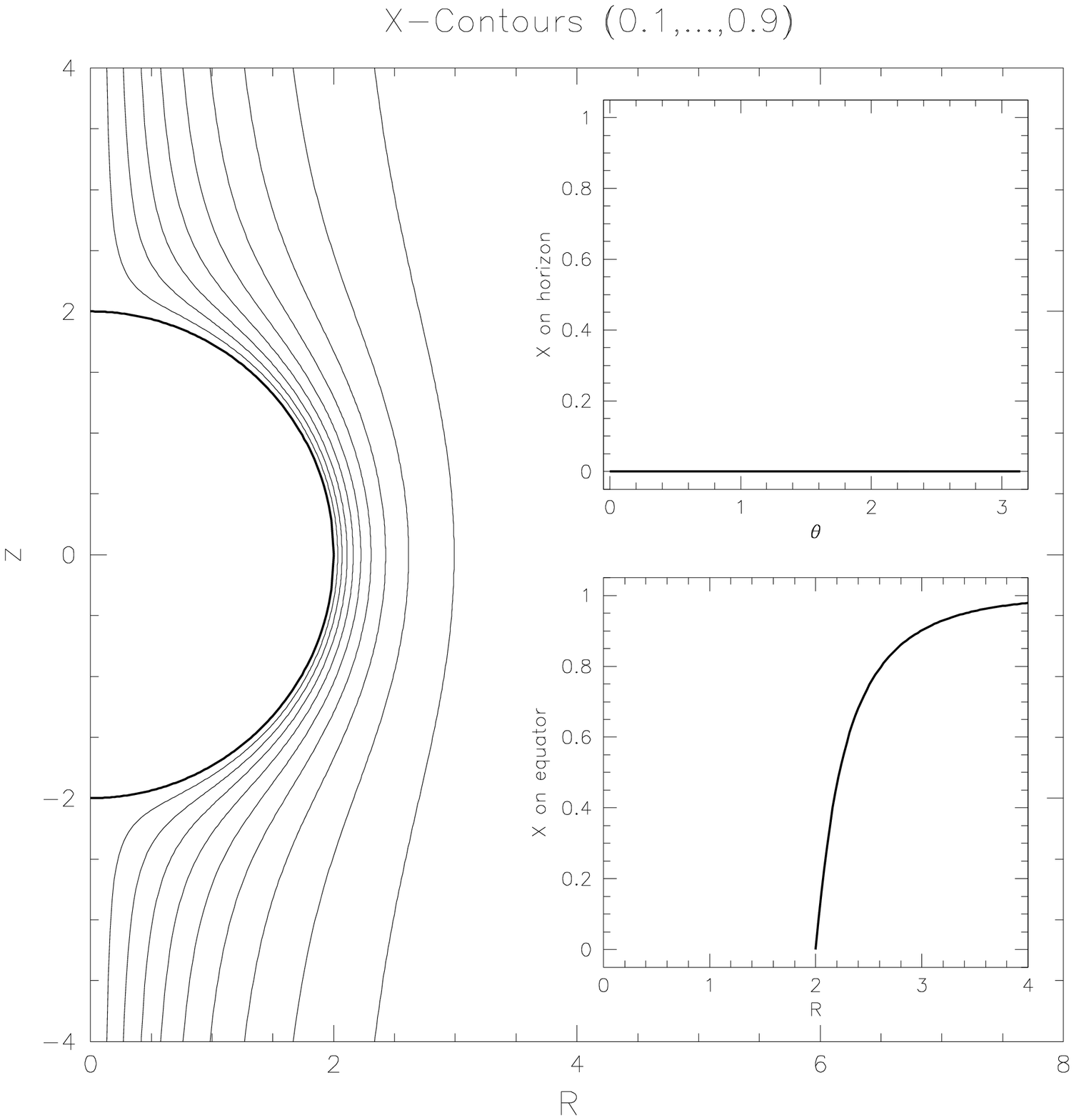}
\end{center}
\caption{\bf As in Fig.7 for the $X$ field
with $N=1$, $\beta=0.05$, $E=1.0$, $q=1.4142 $.}
\label{fig5X}
\end{figure}
\begin{figure}
\begin{center}
\leavevmode
\epsfxsize=440pt
\epsfysize=440pt
\epsfbox{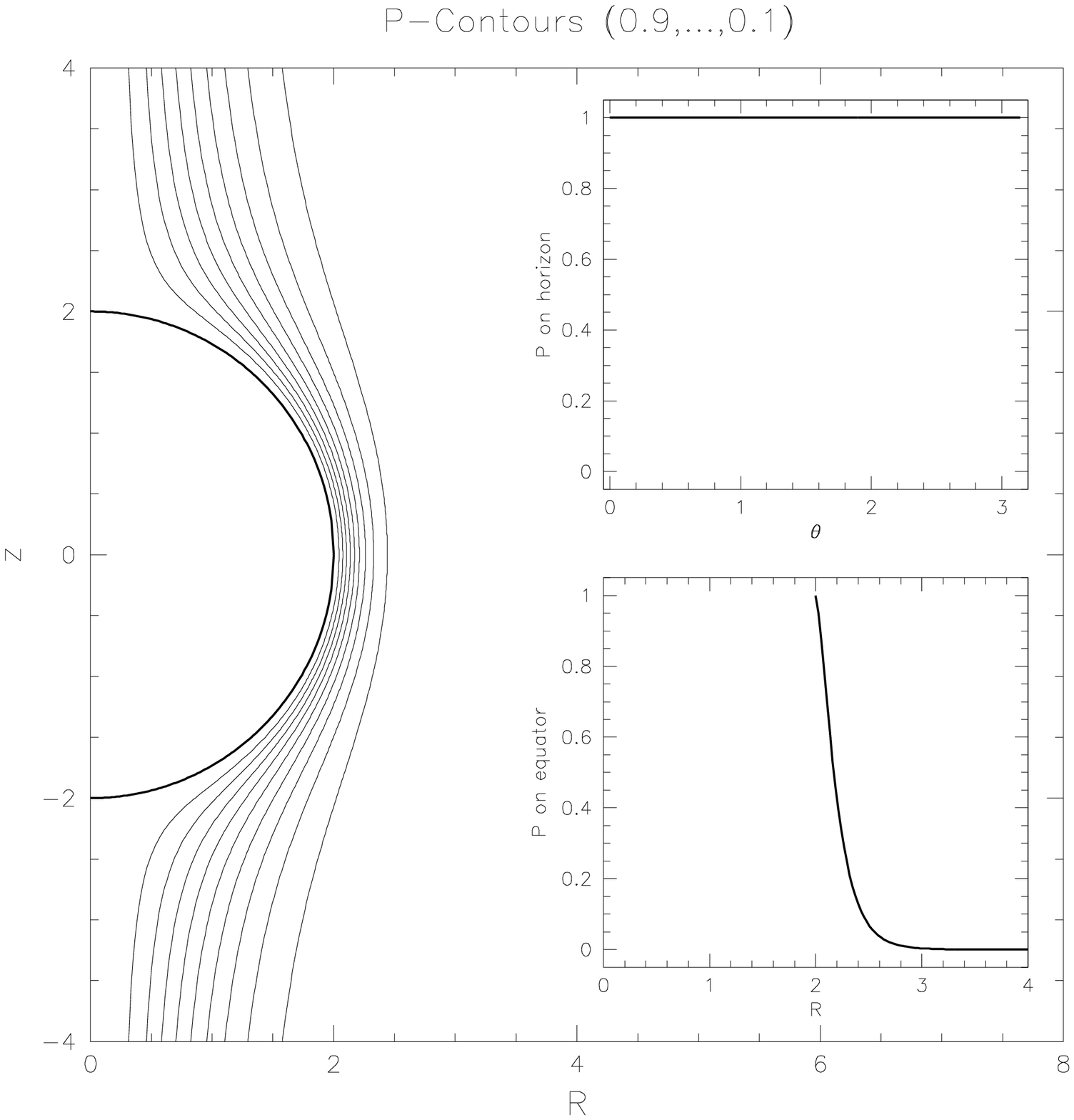}
\end{center}
\caption{\bf As in Fig.8 for the $P$ field
with $N=1$, $\beta=0.05$, $E=1.0$, $q=1.4142$.}
\label{fig5P}
\end{figure}

\begin{figure}
\begin{center}
\leavevmode
\epsfxsize=440pt
\epsfysize=440pt
\epsfbox{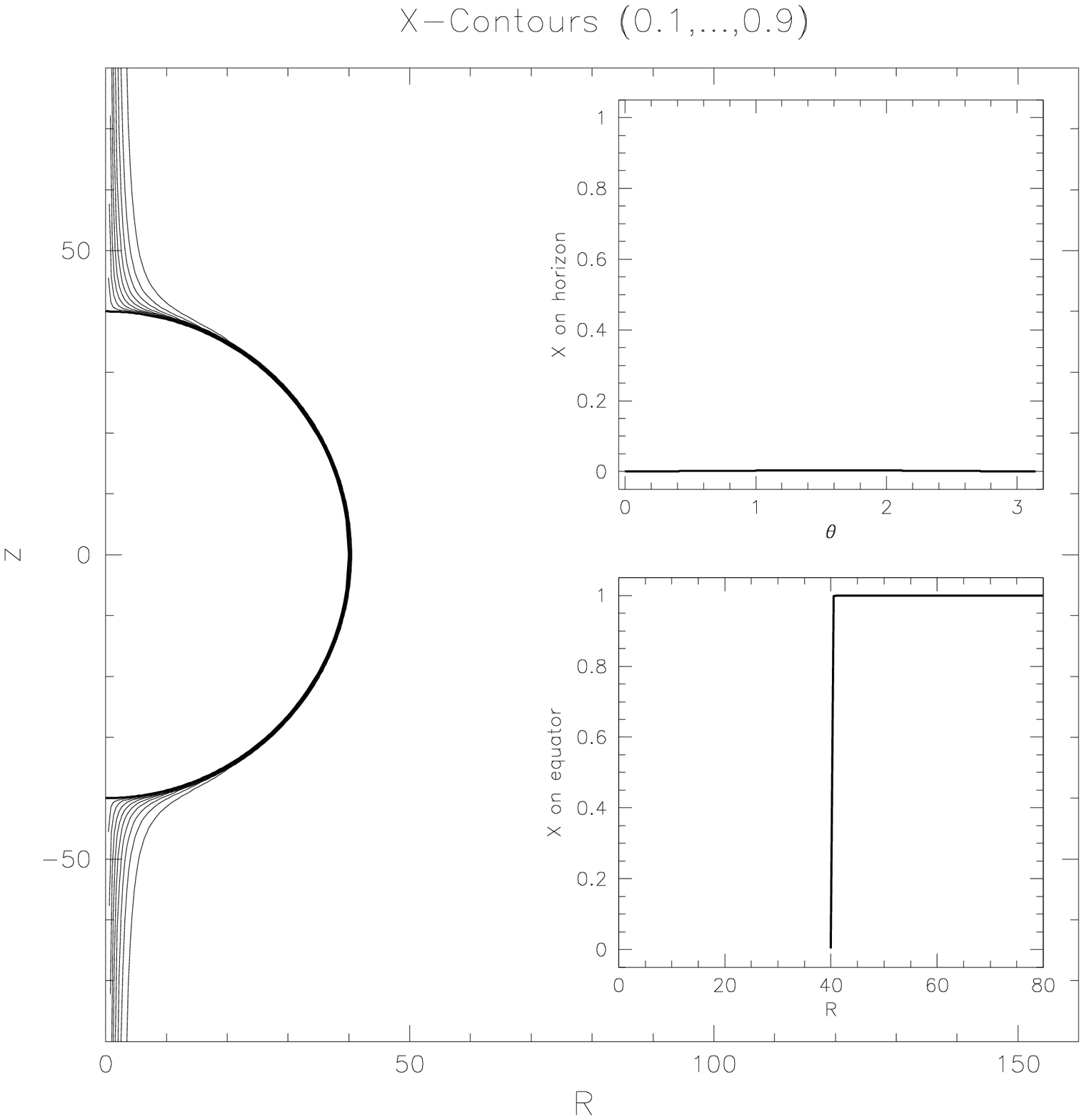}
\end{center}
\caption{\bf As in Fig.9 for the $X$ field
with $N=1$, $\beta=1$, $E=20.0$, $q=28.28$.}
\label{fig8X}
\end{figure}
\begin{figure}
\begin{center}
\leavevmode
\epsfxsize=440pt
\epsfysize=440pt
\epsfbox{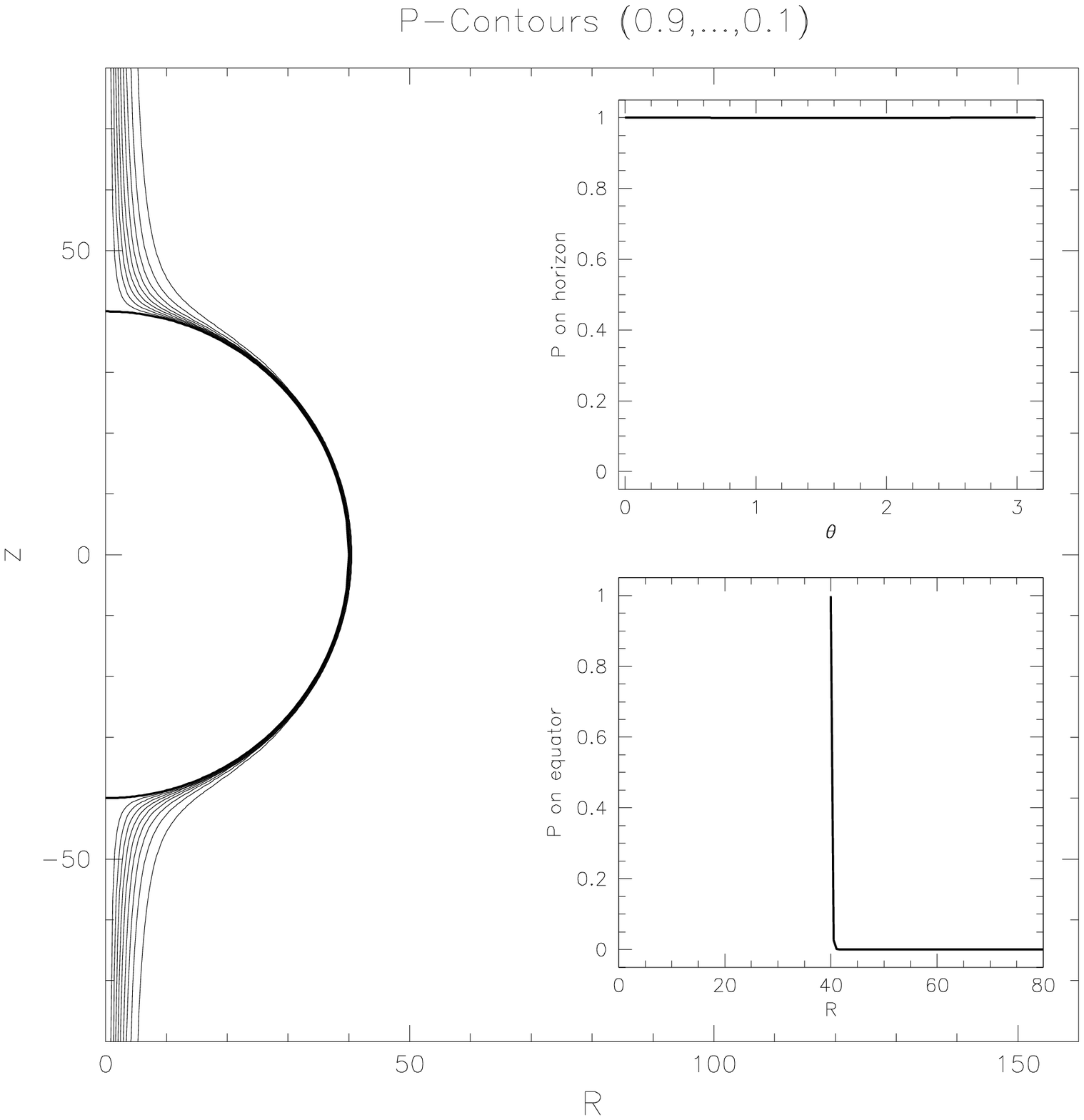}
\end{center}
\caption{\bf As in Fig.10 for the $P$ field
with $N=1$, $\beta=1$, $E=20.0$, $q=28.28$.}
\label{fig8P}
\end{figure}

\end{document}